\documentclass[aps,pre,twocolumn,groupedaddress, showpacs]{revtex4-1}
\usepackage{amsmath,amssymb,graphicx}
\usepackage{scrextend}

\begin{document}


\title{Segment-Scale, Force-Level Theory of Mesoscopic Dynamic Localization and
          Entropic Elasticity in Entangled Chain Polymer Liquids}


\author{Zachary E. Dell}
\affiliation{Department of Physics, University of Illinois, Urbana, IL 61801}

\author{Kenneth S. Schweizer}
\email{kschweiz@illinois.edu}
\affiliation{Department of Materials Science, University of Illinois, Urbana, IL 61801}
\affiliation{Department of Chemistry, University of Illinois, Urbana, IL 61801}
\affiliation{Frederick Seitz Materials Research Laboratory, University of Illinois, Urbana, IL 61801}


\date{\today}

\begin{abstract}
We develop a segment-scale, force-based theory for the breakdown of the unentangled Rouse model and subsequent emergence of isotropic mesoscopic localization and entropic elasticity in chain polymer liquids in the absence of ergodicity-restoring anisotropic reptation motion. The theory is formulated in terms of a conformational $N$-dynamic-order-parameter Generalized Langevin Equation approach. It is implemented using a field-theoretic Gaussian thread model of polymer structure and closed in a universal manner at the level of the chain dynamic second moment matrix. The physical idea is that the isotropic Rouse model fails due to the dynamical emergence of time-persistent intermolecular contacts determined by the combined influence of local chain uncrossability, long range polymer connectivity and a self-consistent treatment of chain motion and the dynamic forces that hinder it. For long chain melts, the mesoscopic localization length (tube diameter) and emergent elasticity predictions are in near quantitative agreement with experiment. Moreover, the onset chain length scales with the semi-dilute crossover concentration with a realistic numerical prefactor. Distinctive predictions are made for various off-diagonal correlation functions that quantify the full spatial structure of the dynamically localized polymer conformation. As the local uncrossability constraint and/or intrachain bonding spring are softened, the tube diameter is predicted to swell until it reaches the chain radius-of-gyration at which point entanglement localization vanishes in a discontinuous manner. A full dynamic phase diagram for the destruction of mesoscopic localization is constructed, which is qualitatively consistent with simulations and the classical concept of an entanglement degree of polymerization.  
\end{abstract}


\maketitle


\section{INTRODUCTION}
\subsection{General Background and Entanglement Phenomenology}
The dynamics and mechanics of concentrated liquids of linear polymers are of broad interest in soft matter science and engineering \cite{2003RubinsteinColby, 1986DoiEdwards, 1979Gennes, ARFM2015LarsonDesai, *RCT2012Roland, PSCR2012Likhtman, COSB2001Ellis, *ARBBS1993ZimmermanMinton, RMP2014BroederszMacKintosh, PR2007RamanathanMorse} . The relevant wide range of length and times scales results in distinctive viscoelastic behavior, the first principles understanding of which is a complex task. A fundamental understanding impacts diverse problems that range from plastics processing  \cite{ARFM2015LarsonDesai, *RCT2012Roland, PSCR2012Likhtman} to biophysical phenomena such as protein folding, chromosome dynamics, and cytoskeleton mechanics \cite{COSB2001Ellis, *ARBBS1993ZimmermanMinton, RMP2014BroederszMacKintosh, PR2007RamanathanMorse, RPP2014HalversonSmrek, *JCP2010BohnHeermann, EL2013LoTurner, *PNAS2016MichielettoTurner, PRL2003GardelValentine}. 
	
Almost all current theories for the dynamics of flexible polymer liquids focus on the motion of a single chain in a sea of identical polymers. The simplest realization, the phenomenological Rouse model \cite{2003RubinsteinColby, 1986DoiEdwards,1979Gennes,JCP1953Rouse-Jr}, coarse grains a polymer to an ideal Gaussian random walk composed of $N$ statistical segments (size, $\sigma \sim$ nm) linearly connected by entropic harmonic springs with radius-of-gyration $R_g^2 = N \sigma^2/6$. All consequences of intermolecular forces are empirically modeled as a frictional drag force on each segment and a corresponding white noise random force. Though useful for the generic dynamics of short chain melts, Rouse theory breaks down in solutions due to hydrodynamics \cite{2003RubinsteinColby, 1986DoiEdwards, JCP1956Zimm}, near the glass transition due to nonlocal viscoelastic effects \cite{ARFM2015LarsonDesai, *RCT2012Roland,ACR2011Freed,JPCL2013MirigianSchweizer, M2015MirigianSchweizer}, and for long chain liquids due to ``entanglements'' \cite{2003RubinsteinColby, 1986DoiEdwards, 1979Gennes, ARFM2015LarsonDesai, *RCT2012Roland, PSCR2012Likhtman, PPS1967Edwards,JCP1971Gennes, AP2002McLeish}. The focus of this article is the dynamic emergence of the latter, and implications for spatial localization and entropic shear rigidity.  

Perhaps the most spectacular consequence of entanglements is the intermediate time rubbery plateau of the dynamic shear stress relaxation modulus in the liquid state. Motivated by an analogy to crosslinked rubbers, deGennes \cite{1979Gennes,JCP1971Gennes} postulated it arises from dynamical constraints of surrounding polymers on a tagged chain for sufficiently large amplitude motions resulting in transient localization in an Edwards \cite{1986DoiEdwards,PPS1967Edwards} tube-shaped region of space. Long time relaxation, flow and diffusion occur via a quasi-1D stochastic motion, reptation \cite{1979Gennes,JCP1971Gennes} (Figure 1). A critical element is an adjustable parameter: the tube diameter, $d_T$, or average transverse dynamic localization length. In the large $N$ limit it depends only on polymer chemistry and concentration \cite{2003RubinsteinColby,M1994FettersLohse}. 
	
The rubber analogy suggests a mean number of $N_e \equiv d_T^2/\sigma^2$  segments between two entanglements or \textit{effective} crosslinks with a corresponding modulus of \cite{2003RubinsteinColby, 1986DoiEdwards, 1979Gennes, M1994FettersLohse, S2004EveraersSukumaran, PRL1987KavassalisNoolandi, *M1987Lin, PY1989WittenMilner}:
\begin{equation}
G_e = C \frac{\rho_s k_BT}{N_e} = C \frac{\rho_s \sigma^2 k_BT}{d_T^2} = \tilde{C} \frac{k_BT}{p^3},
\label{eq1}
\end{equation}
where $\rho_s$ is the segmental number density, $k_BT$ is thermal energy, and $C$ is a constant of order unity. The final equality ($\tilde{C} \approx 2.3\times10^{-3}$ \cite{M1994FettersLohse}) connects entanglement density to coil geometric interpenetration via the ``packing length'' \cite{M1994FettersLohse, S2004EveraersSukumaran, PRL1987KavassalisNoolandi, *M1987Lin, PY1989WittenMilner} which is the ratio of the space-filling polymer volume to its mean squared end-to-end distance. In melts, $p = (\rho_s \sigma^2)^{-1} \approx 0.15 - 0.5$ nm, and it determines the mesoscopic tube diameter as $d_T = 18p \approx 3-10$ nm \cite{M1994FettersLohse}. The critical degree of overlap for one entanglement is when $\sim 20$ chains inhabit a region of space of order a polymer coil volume  \cite{2003RubinsteinColby, S2004EveraersSukumaran, PRL1987KavassalisNoolandi, *M1987Lin, PY1989WittenMilner}. 

In solutions, the geometric overlap density (dilute-semidilute crossover \cite{2003RubinsteinColby, 1986DoiEdwards, 1979Gennes}) obeys $\rho_p^* 4\pi R_g^3/3 \approx 1$, where $\rho_p=\rho_s/N$. The entanglement onset is quantitatively higher, $\rho_{0, s} \propto \rho_s^* = \left( 9\sqrt{6} /2\pi \right) N^{-1/2}\sigma^{-3} $; equivalently, the onset chain length $N_{on} \propto \rho_s^{-2} \propto p^2$. Polymers in dilute good solvents are swollen, $R_g \propto N^{3/5}$, so $\rho_{0,s} \propto \rho_s^* \propto N^{-4/5}$ and the other scaling relations change accordingly \cite{2003RubinsteinColby, 1986DoiEdwards, 1979Gennes}. 
	
Despite its successes in equilibrium, the reptation-tube model does not address the most fundamental question: \textit{why and how does the isotropic Rouse model qualitatively fail when chains get sufficiently long and/or concentrated}? This is our focus, and the answer constitutes what we mean by ``emergence of dynamic entanglements and a tube''. Little theoretical work has been done to address this difficult issue. For context, we briefly summarize prior attempts germane to the questions we address. 

\subsection{Previous Microscopic Theories and Our Approach}
Existing ``microscopic'' dynamical theories of entangled polymers fall into two broad categories: topological approaches and Generalized Langevin Equation (GLE)-based theories. The most developed former type of approach treats interactions asymmetrically: equilibrium structure is per an ideal gas of hypothetical molecules of zero volume, and dynamic uncrossability is strictly enforced at a 2-body level and approximately at higher levels \cite{PRL1993Szamel, JCP1994SzamelSchweizer, PRL2011SussmanSchweizer, PR2011SussmanSchweizer, PRL2012SussmanSchweizer, JCP2013SussmanSchweizer, SM2015KimDutta}. In GLE-type  \cite{JCP1989Schweizer1, *JCP1989Schweizer2, MTS1997SchweizerFuchs, 2004KimmichFatkullin, JCP1999Guenza, PR2014Guenza} approaches, polymers experience nonlocal in space and time effective forces (memory functions) where dynamic uncrossability and equilibrium packing in the liquid are determined by the same interaction potentials. 
 
 \begin{figure}[h!]
 	\includegraphics[width = \columnwidth]{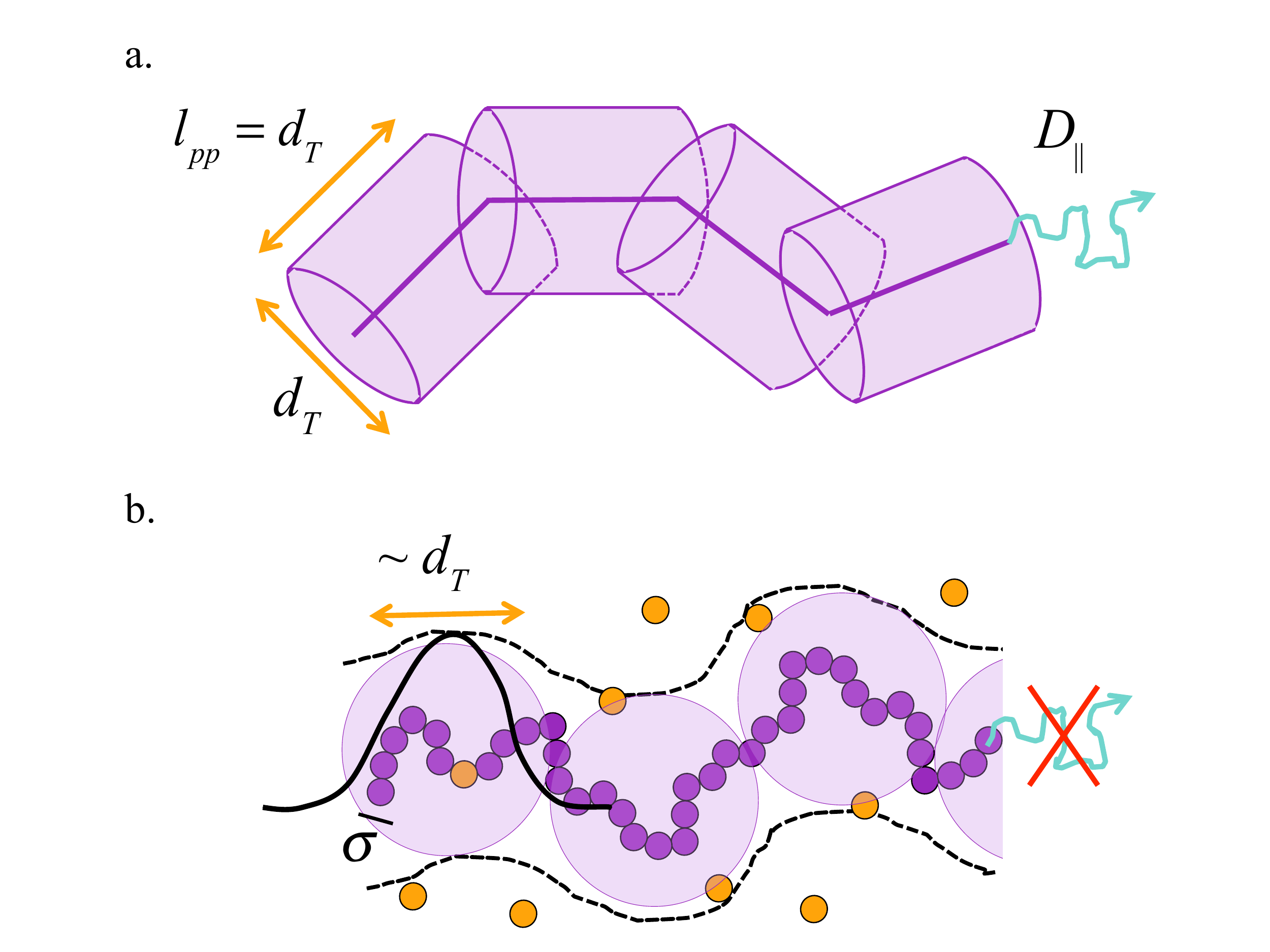}%
	\caption{a) The dynamic topological theory of \cite{PRL2012SussmanSchweizer, JCP2013SussmanSchweizer} a priori coarse-grains polymer chains to the primitive path (PP) level. The PP length is self-consistently computed using $l_{pp} = d_T$, which characterizes transverse localization. Segmental degrees of freedom inside the PP scale are ignored.  b) Our work focuses on the breakdown of the isotropic Rouse model analyzed at the more microscopic segment scale, $\sigma$. Anisotropic reptation is not allowed, and segments can localize in a spherical region of diameter $d_T$.\label{fig1}}
\end{figure}

The primary topological approach was initiated by Szamel and is built on a Smoulchowski description for a gas of non-rotating, infinitely thin, dynamically uncrossable rigid needles; tube localization and reptation dynamics are predicted \cite{PRL1993Szamel,JCP1994SzamelSchweizer,PRL2011SussmanSchweizer,PR2011SussmanSchweizer}. To treat flexible chains, the polymer is a priori coarse-grained at the uncrossable needle primitive path (PP) level where $l_{pp} = d_T$  (see Fig.1a)\cite{PRL2012SussmanSchweizer, JCP2013SussmanSchweizer}. To predict tube localization, the PPÕs are disconnected, and their size is self-consistently computed yielding a reasonable value of $d_T \approx 10 p$ \cite{PRL2012SussmanSchweizer, JCP2013SussmanSchweizer}. However, beyond the neglect of PP rotation and connectivity, all sub-PP (segmental scale) dynamic fluctuations are ignored. Such an approach is not fully ``bottom-up'', and does not address dynamic correlations inside, or well beyond, the tube scale. Recently, an alternative topological approach was proposed based on an analogy to phenomenological models of superconductivity \cite{SM2015KimDutta}.

Predictive microscopic GLE theories at the isotropic single chain dynamics level were formulated long ago by Schweizer in two ways: the renormalized Rouse (RR) model and the polymer mode coupling theory (PMCT) \cite{JCP1989Schweizer1, *JCP1989Schweizer2, MTS1997SchweizerFuchs, 2004KimmichFatkullin}. Both relate slow entangled dynamics to structure. The primary focus was scaling of relaxation times and transport coefficients, and anomalous diffusion at intermediate times, issues \textit{not} of present interest. Two other key differences are (i) PMCT and RR theories did \textit{not} perform a non-perturbative, self-consistent calculation of confining forces and chain motion, and (ii) $N$-dependent renormalizations were argued to be conceptually related to the long range deGennes correlation hole \cite{1979Gennes}. These aspects are \textit{not} adopted here. Although our starting point remains a set of coupled single chain GLEs, we propose a self-consistent theory that is closed at the level of the connected chain second moment which involves $N$ coupled conformational dynamic order parameters. Uncrossability enters \textit{only} locally. The key is chain connectivity in conjunction with self-consistency between polymer motion and slowly relaxing interchain forces (persistent contacts) on all length scales. 
 
Our goals also differ from prior GLE-based efforts since we seek to understand the origin of the breakdown of isotropic Rouse dynamics in the absence of any ergodicity-restoring (e.g., reptation) motion (Fig.1b). We believe the fundamental reason that a long chain moves at long times via anisotropic reptation can be understood as due to its inability to exploit three spatial dimensions. Thus, understanding the physics of isotropic localization can provide an objective justification for \textit{why} the \textit{dynamical} effect called ``entanglement'' necessitates anisotropic transport. Interestingly, liquids of cyclic ring polymers (no free ends) cannot employ reptation for long time/distance relaxation \cite{RPP2014HalversonSmrek, *JCP2010BohnHeermann}. Although there are structural complexities in ring melts not present in their chain analog, it is fascinating to consider the possibility that literal mesoscopic localization can occur in them. Recent simulation studies  \cite{EL2013LoTurner, *PNAS2016MichielettoTurner} have suggested a kinetically arrested mesoscopic ``topological glass'' can emerge in ring melts at large enough molecular weights. In a general sense, our present work may be relevant to this problem. 
 
The theoretical tools we employ have a long history of success for identifying the \textit{dynamic crossover} to a much slower motional mechanism in simpler systems, e.g., persistent caging in simple liquids as a predictive indication of the crossover to glassy activated dynamics  \cite{JPCL2013MirigianSchweizer, JSMTE2005ReichmanCharbonneau, 2008Gotze}. We emphasize our ability to compute the full spatial structure of the dynamic confinement field. Emergent entropic rigidity follows immediately, along with other spatially-resolved correlation functions. We explicitly demonstrate that softening dynamical uncrossability constraints in our theory leads to the destruction of mesoscopic localization, consistent with polymer simulations \cite{PSCR2012Likhtman, PRL1991DueringKremer, M2014KalathiKumar, PNAS2015ZhangWolynes}. 

The fact that we analyze localization in an isotropic framework is, we believe, only a quantitative issue. As concrete support for this view, we note that Sussman and Schweizer  \cite{PRL2011SussmanSchweizer, JCP2013SussmanSchweizer} have shown that for needle fluids where dynamic uncrossability \textit{is} exactly enforced at the two polymer level, mesoscopic localization is predicted in an almost quantitatively identical manner regardless of whether needles are allowed to move in a 3D isotropic manner or anisotropically with reptation quenched and localization only in 2 transverse directions. Finally, we mention the ``many chain'' approaches of Guenza \cite{JCP1999Guenza,PR2014Guenza} which relate correlation hole structure and slow cooperative dynamics of interpenetrating chains. This work is very different than ours, emphasizes dynamics, and does not a priori address entanglement localization. 

Section II formulates the self-consistent GLE theory of dynamic localization. Two simplified limits are discussed in Section III. The universal Gaussian thread model of chain liquid structure is adopted in Section IV to construct the specific dynamical theory we analyze here. Section V presents our primary results for chain melts and semi-dilute solutions. What the theory predicts if dynamical uncrossability constraints are softened is studied in Section VI, and a ``disentanglement phase diagram'' is constructed. The paper concludes in Section VII with a discussion. Technical details and additional results are contained in the Appendices. Appendix I addresses general theory issues, Appendix II the two limiting cases, and Appendix III derives analytic results in the thread polymer limit.

\section{GENERAL THEORY}	
\subsection{Rouse Model}
	The Rouse model consists of $N$ overdamped Langevin equations for the positions of chain segments (size $\sigma$). If $\vec{R}_\alpha$  is the position of segment $\alpha$, ignoring end effects one has \cite{1986DoiEdwards, JCP1953Rouse-Jr}:
\begin{equation}
\zeta_s \frac{d}{dt} \vec{R}_\alpha(t) = k_s \left[ \vec{R}_{\alpha+1}(t) - 2 \vec{R}_\alpha (t) + \vec{R}_{\alpha -1}(t) \right] + \vec{\xi}_\alpha(t)
\label{eq2}
\end{equation}
 where $\zeta_s$ is the segmental friction constant, $k_s = 3k_BT/\sigma^2$ is the entropic spring constant, \linebreak $\sigma^2 = \left< \left(\vec{R}_\alpha - \vec{R}_{\alpha-1}\right)^2\right>$, and $R_g^2 = N\sigma^2/6$. The fluctuating random forces obey:
\begin{equation}
\left< \vec{\xi}_\alpha(t) \cdot \vec{\xi}_\beta(t') \right> = 6k_BT \zeta_s \delta_{\alpha \beta} \delta(t-t')
\label{eq3}
\end{equation}
which characterizes the main dynamical assumption of the Rouse model--forces on segments of a tagged polymer due to other chains are uncorrelated in space and time.  

Equations (2) can be decoupled by introducing normal (Rouse) modes \cite{1986DoiEdwards, JCP1953Rouse-Jr}:
\begin{equation}
\vec{R}_\alpha(t) = \sum_{p=0}^N \vec{X}_p(t) \psi_p(\alpha)
\label{eq4}
\end{equation}
where $\vec{X}_p(t)$ is the time-dependent mode amplitude, and:
\begin{equation}
\psi_p(\alpha) = \left \{ 
\begin{array} {cc}
N^{-1/2} & p=0\\
(2/N)^{1/2} \cos(p\pi\alpha/N) & p\neq 0
\end{array}
\right .
\label{eq5}
\end{equation}
The $p=0$ mode describes the chain center-of-mass (COM) motion, and $p\neq 0$ modes represent internal conformational fluctuations on a length scale $\Lambda_p = \sigma \sqrt{N/p}$. Using Eqs. (4) and (5) in Eq. (2) yields the mode amplitude equations of motion:
\begin{equation}
\zeta_s \frac{d}{dt} \vec{X}_p(t) = -\lambda_p \vec{X}_p(t) + \vec{\xi}_p(t)
\label{eq6}
\end{equation}
where $\lambda_p = p^2 \pi^2 k_s/N^2$ is the mode spring constant and $\vec{\xi}_p$ the corresponding random fluctuating force. The equilibrium mode amplitude correlations from Eq. (6) are:
\begin{equation}
\left< X_{p, i} X_{q, j} \right> =  \delta_{pq} \delta_{ij} \frac{k_BT}{\lambda_p}
\label{eq7}
\end{equation}
where $i, j$ indicate Cartesian components. All higher correlations can be expressed in terms of Eq. (7) due to the Gaussian nature of Rouse theory. From Eqs. (4)-(7) the chain correlations can be expressed in terms of the mode correlations, yielding: 	
\begin{align}
\left< \left|\vec{R}_\alpha(t) - \vec{R}_\gamma(0)\right|^2 \right> &= \sum_{p=0}^N \left( \psi_p(\alpha) - \psi_p(\gamma)\right)^2 \nonumber \\
&+\sum_{p=0}^N \psi_p(\alpha)\psi_p(\gamma) \left< \Delta \vec{X}_p^2(t) \right>
\label{eq8}
\end{align}
where $\Delta \vec{X}_p(t) \equiv \vec{X}_p(t) - \vec{X}_p(0)$. The first term represents the equilibrium correlations where $\left< \left|\vec{R}_\alpha- \vec{R}_\gamma\right|^2 \right> = \left | \alpha - \gamma \right | \sigma^2$ , and the second term can be determined from Eq. (6).  

 \begin{figure*}[t!]
 	\includegraphics[height = 2.5in]{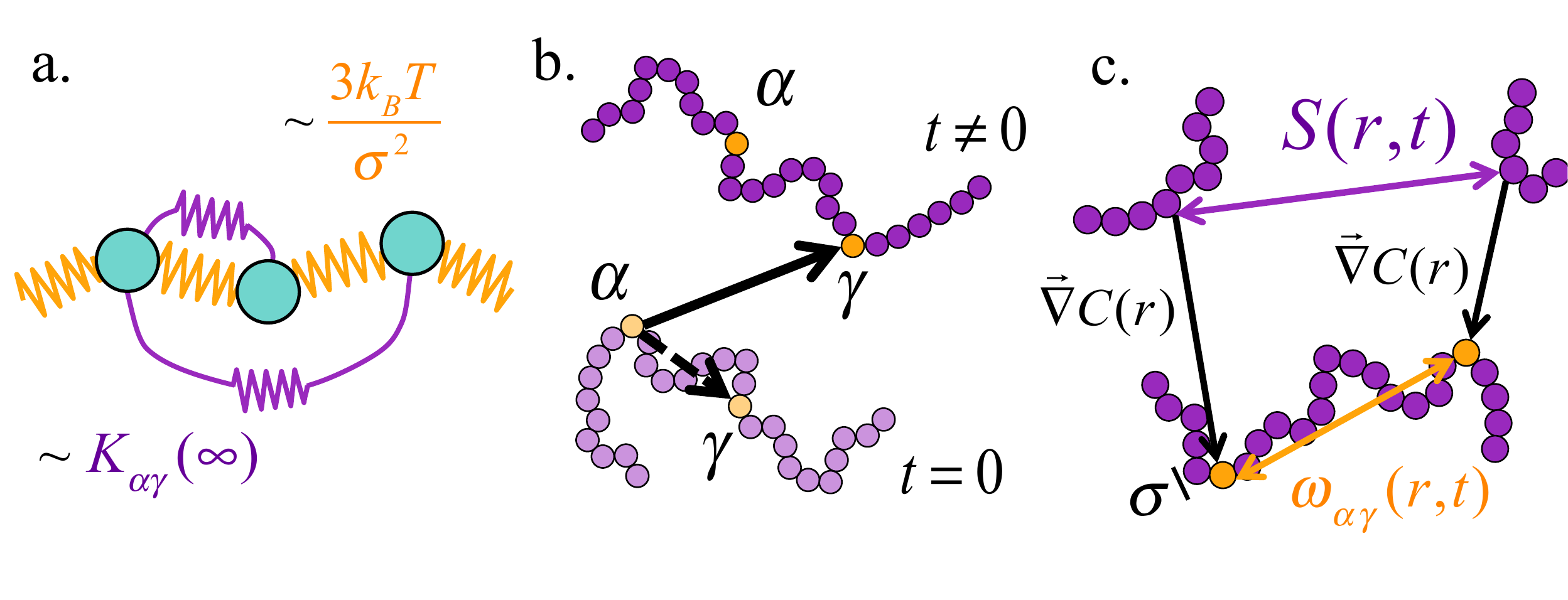}%
	\caption{Theoretical overview. a) In the long time kinetically arrested state the generalized Rouse model has two sets of springs: bonded intra-chain (orange) entropic springs and dynamically emergent inter-chain (purple) springs between all segments determined by the self-consistently calculated arrested dynamic force correlation function matrix, $K_{\alpha \gamma}(t\rightarrow\infty)$. b) The theory is closed at the level of the chain second moment where the solid and dashed arrows indicate the two terms in Eq.(15). c) Interchain force-force time correlations involve an effective forces, $k_BT\vec{\nabla}C(r)$, and their dynamical relaxation is due to tagged chain motion, via $\omega_{\alpha \gamma}(r,t)$, and motion of the surrounding polymers, via $S(r,t)$.\label{fig2}}
\end{figure*}

\subsection{Kinetically Arrested Generalized Rouse Model} 
	Our starting point is the formally linear, coupled, nonlocal in space and time, GLE equations-of-motion for a tagged chain which are easily derived using standard Mori-Zwanzig projection operator methods to be \cite{JCP1989Schweizer1, *JCP1989Schweizer2, JCP1974Zwanzig}:
\begin{align}
\zeta_s \frac{d}{dt} \vec{R}_\alpha(t) &= k_s \left[ \vec{R}_{\alpha+1}(t) - 2 \vec{R}_\alpha (t) + \vec{R}_{\alpha -1}(t) \right] + \vec{\xi}_\alpha(t) \nonumber \\
& -\int_0^t d\tau \sum_{\gamma = 1}^N K_{\alpha \gamma}(t-\tau) \frac{d\vec{R}_\gamma(\tau)}{d\tau} + \vec{F}_\alpha(t)
\label{eq9}
\end{align}
The first three terms are identical to the Rouse model. The last two terms are due to the more slowly relaxing forces beyond the $\sigma$ scale, meant to capture nonlocal viscoelastic effects. Here, $\vec{F}_\alpha$ is the (formally projected \cite{JCP1974Zwanzig}) slowly relaxing component of the total force on segment $\alpha$ from surrounding chains and the memory function matrix $K_{\alpha\gamma}$ is:
\begin{equation}
\left< \vec{F}_\alpha(t) \cdot \vec{F}_\gamma(t') \right> = 3k_BT K_{\alpha \gamma}(t-t')
\label{eq10}
\end{equation}
Ignoring chain end effects implies that any two segment correlations can be expressed in terms of their arc-length separation $\Delta \alpha \equiv \left |  \alpha - \gamma \right |$, and thus $K_{\alpha \gamma} \rightarrow K(\Delta \alpha)$. Eq. (9) can thus be diagonalized by Rouse modes to yield independent mode amplitude equations: 
\begin{align}
\zeta_s \frac{d}{dt} \vec{X}_p(t) &= -\lambda_p \vec{X}_p(t) + \vec{\xi}_p(t) \nonumber \\
&  - \int_0^t d\tau \,K_p(t-\tau) \frac{d\vec{X}_p(\tau)}{d\tau}  + \vec{F}_p(t) \label{eq11} \\
K_p(t) &= \frac{1}{A_p} \sum_{\Delta \alpha= 0}^N \psi_p(\Delta \alpha) K(\Delta \alpha, t)
\label{eq12}
\end{align}
where $A_0 = N^{-1/2}$ and $A_{p\neq0} = (2/N)^{1/2}$.

If long time force correlations decay to zero (liquid), the memory function contribution in Eq.(11) enters as an effective friction constant matrix. If force correlations persist at long times, emergent solid-like behavior is predicted. For this case of interest, the memory function matrix reduces to effective springs (Fig. 2a) in Eqs. (9) and (11):  
\begin{align}
\zeta_s \frac{d}{dt} \vec{X}_p(t) &= -\lambda_p \vec{X}_p(t) + \vec{\xi}_p(t) \nonumber \\
&  -K_p(\infty) \left(\vec{X}_p(t) - \vec{X}_p(0) \right)  + \vec{F}_p(t) \label{eq13} 
\end{align}
The effective springs are coupled since $K_p(\infty)$ depends on length scale, or equivalently mode index, $p$. It describes in real space a set of springs that connect a tagged segment to \textit{every other} segment on the chain (Fig.2a). This might seem reminiscent of phenomenological slip-link or slip-spring models \cite{M2005Likhtman,*M2008ReadJagannathan, ARCBE2014SchieberAndreev, *M2006NairSchieber, *PRL2008KhaliullinSchieber}, but it is not since we do not formulate our approach at the coarse-grained tube diameter or PP level, springs connect segments on all scales, and we develop a microscopic theory to self-consistently compute the effective springs. Overall, Eq.(13) corresponds to a description of a kinetically arrested polymer with two types of springs: dynamically emergent matrix of springs determined by \textit{inter}-molecular interactions and intramolecular bonded entropic Rouse springs. At long times, the mode correlations follow from Eq. (13) as:	
\begin{align}
\left< \Delta \vec{X}_p^2(\infty) \right> &= \frac{3k_BT}{(p\pi/N)^2 + 2 K_p(\infty)}
\label{eq14}
\end{align}
 Finally, the dynamic second moment of the chain in real space is:
\begin{align}
\delta \mu_{\alpha \gamma}^{(2)} (\infty) &= \left< \left|\vec{R}_\alpha(\infty) - \vec{R}_\gamma(0)\right|^2 \right> -  \left< \left|\vec{R}_\alpha(0) - \vec{R}_\gamma(0)\right|^2 \right> \nonumber \\
&= \frac{3k_BT}{2K_0(\infty)} \frac{1}{N} + \sum_{p=1}^N \frac{\psi_p(\Delta\alpha)}{\sqrt{2N}}\frac{3k_BT}{(p\pi/N)^2 + 2 K_p(\infty)}
\label{eq15}
\end{align}
The first line defines the dynamic portion of the chain second moment, $\delta \mu_{\alpha \gamma}^{(2)}$ (see Fig. 2b) as the difference between the full second moment of the chain and the equilibrium contribution. The second line follows from Eq.(8) by ignoring end effects. The $p=0$ COM contribution is separated from the internal mode contributions. If long time force correlations persist, Eq.(15) can potentially (\textit{not} guaranteed) predict a nonzero localization length (taken as the estimated tube diameter) deduced from the $\alpha = \gamma$ term:
\begin{align}
\delta \mu_{\alpha \alpha}^{(2)} (\infty) \equiv r_{loc}^2 \equiv \frac{d_T^2}{4}.
\label{eq16}
\end{align}
Equation (15) does not depend on the specific model of the force correlations, and a self-consistent theory for their slowly relaxing components must be constructed.  

\subsection{Effective Force Theory and Dynamic Closure}
	Our theory for the force-force correlation (memory) function matrix is developed in Appendix I. We consistently invoke a Gaussian density fluctuation perspective for both equilibrium and dynamic correlations, and construct a closed theory at the single chain level that has $N$ dynamic order parameters. The 3 key simplifications are as follows. (i) Interchain segment-segment forces are replaced by effective forces determined by pair structure which are \textit{spatially local} if the direct intermolecular interactions are short range (the case of present interest). (ii) Relaxation of force correlations proceeds in parallel via tagged single chain relaxation and collective matrix density relaxation, which are directly related in the arrested ÒentangledÓ state. (iii) Projected dynamics is replaced by its real Newtonian analog and full self-consistency between tagged chain motion and force time correlations is enforced. These ideas lead to:
\begin{align}
K_{\alpha \gamma}(t) = \frac{\beta^{-1} \rho_s}{3} \int \frac{d\vec{k}}{(2\pi)^3} \left( kC(k) \right)^2 \omega_{\alpha\gamma}(k,t) S(k,t). 
\label{eq17}
\end{align}
Here $\rho_s$ is the segmental number density, $C(k)$ the Fourier space interchain segment-segment direct correlation function\cite{1986HansenMcDonald, ACP1997SchweizerCurro}, $h(k)$ the non-random part of the segment-segment pair distribution function, $h(r) = 1-g(r)$, and $S(k) = \omega(k) + \rho_s h(k)$ is the collective density fluctuation static structure factor. Dynamic correlations decay via single chain and collective liquid motions, as statistically quantified by $\omega_{\alpha \gamma}(k,t)$ and $S(k,t)$, respectively, where:
\begin{align}
\omega_{\alpha \gamma}(k, t) &\equiv \left< \exp \left[ -i \vec{k} \cdot \left( \vec{R}_\alpha(t) - \vec{R}_\gamma(0)\right)\right]\right>  \nonumber \\
&\qquad \xrightarrow[t=0]{} e^{-k^2 \sigma^2\left | \alpha - \gamma\right|/6} . 
\label{eq18} \\
\omega(k,t) & = \frac{1}{N} \sum_{\alpha, \gamma = 1}^N \omega_{\alpha \gamma}(k,t)
\label{eq19}
\end{align}
and  $S(k,t)$ is defined analogous to Eq. (19) but averaged over all segments in the liquid. 

A schematic of the real space interpretation of Eq. (17) is shown in Fig. 2c. Effective forces are $\vec{F}_{eff} (r) = k_BT \vec{\nabla} C(r)$; for polymers that repel via hard-core-like interactions, they capture purely \textit{local} excluded volume effects which set the strength of uncrossability forces. At $t=0$, Eq. (17) sums up dynamical constraints on different length scales due to interactions separated in space and time which are correlated via chain connectivity. Their temporal persistence is described by the two time dependent functions in Eq. (17), and to close the theory requires explicit expressions for them. 	

Starting with Eq.(18), a 2nd order cumulant expansion per the Gaussian idea gives: 
\begin{align}
\omega_{\alpha\gamma}(k,t) & \approx \exp\left[ - \frac{k^2 \mu_{\alpha \gamma}^{(2)}(t)}{6}\right]
\label{eq20}
\end{align}
or, 
\begin{align}
\omega_{\alpha\gamma}(k,t) & \approx \exp\left[ - \frac{k^2 \sigma^2\left | \alpha - \gamma \right |}{6}\right] \exp\left[ - \frac{k^2 \delta\mu_{\alpha \gamma}^{(2)}(t)}{6}\right]
\label{eq21}
\end{align}
where $\delta \mu^{(2)}_{\alpha\gamma}(t)$ is the time dependent contribution in Eq. (15). The collective propagator is defined as $\Gamma(k,t) \equiv S(k,t)/S(k)$. Idea (ii) above is invoked to determine it, i.e., for entanglement-induced arrest of isotropic motion, collective dynamics is slaved to single chain dynamics. This self-consistent dynamic mean-field-like simplification closes the theory at the single chain level. In simple liquids, it is called a Vineyard approximation \cite{1986HansenMcDonald, PR1958Vineyard}. For polymers, two limiting implementations correspond to assuming (i) segmental or (ii) coherent chain motions of the surrounding polymer matrix are necessary to relax intermolecular forces. The ''segmental Vineyard'' closure approximation is:
\begin{align}
\Gamma(k,t) \approx \Gamma_s(k,t) = \exp\left[ - \frac{k^2 \delta\mu_{\alpha \alpha}^{(2)}(t)}{6}\right]
\label{eq22}
\end{align}
Given the neglect of end effects, $ \delta\mu_{\alpha \alpha}^{(2)}$ is independent of $\alpha$. In the long time limit, Eq. (22) then yields a localized Gaussian Debye-Waller form, which from Eq. (16) is:
\begin{align}
 \Gamma_s(k,t) = \exp\left[ - \frac{k^2 d_T^2}{24}\right],
\label{eq23}
\end{align}
The ``chain Vineyard'' closure approximation is:
\begin{align}
\Gamma(k,t) \approx \Gamma_c(k,t) = \frac{\omega(k,t)}{\omega(k)}
\label{eq24}
\end{align}
A priori, it is not obvious which of these two closure approximations is ``more rigorous'' or ``better''. Reassuringly, we will show that they yield qualitatively (almost quantitatively) identical results. Hence, for the remainder of this work we focus mainly on the simpler Eq. (23). Comments on differences will be made when appropriate.  
	
The long-time arrested limits of Eqs. (17) and (21) with Eq. (23) or (24) fully define our theory based on $N$ coupled GLEs. Ignoring chain end effects implies that $\mu^{(2)}_{\alpha \gamma}$, $\omega_{\alpha \gamma}$, and $K_{\alpha \gamma}$  depend on $\left | \alpha - \gamma \right |$ alone. Hence, the analysis of Sec.II.B applies, resulting in a self-consistent closure for $\delta\mu^{(2)}_{\alpha \gamma}(\infty)$ via $K_p(\infty)$ in Eq.(15). Physically, the self-consistency means that spatially-resolved force relaxation depends on chain dynamics, but chain dynamics are determined by force relaxation. This set of equations is the fundamental result of the paper. The equilibrium pair structure is required as input.  

\section{SIMPLIFIED VERSIONS OF THE DYNAMIC THEORY}
Two simplified limits of our general theory are glassy localization on the segmental scale, and a center-of-mass (COM) or long wavelength limit for mesoscopic localization. Both avoid the $N$ coupled dynamic order parameters aspect of the full theory, and close the theory for the segmental localization length per Eqs.(15) and (16).
\subsection{Glassy Diagonal Limit}
 	The most na\"{i}ve approach introduces a viscoelastic memory diagonal in segment coordinates: $K_{\alpha \gamma}(t) \approx K_{\alpha \alpha}(t) \cdot \delta_{\alpha \gamma}$. This approximation can be viewed as retaining only the most local $p=N$ part of the memory function. Such a diagonal approximation effectively \textit{disconnects} the segments \textit{in} the memory function since $\omega(k,t) \rightarrow \Gamma_s(k,t)$, where $\Gamma_s$ is given by Eq. (23), thereby yielding: 
\begin{align}
K_{Diag}(\infty) = \frac{\beta^{-1} \rho_s}{3} \int \frac{d\vec{k}}{(2\pi)^3} \left( kC(k) \right)^2 S(k) \exp\left(- \frac{k^2d_T^2}{12} \right) 
\label{eq25}
\end{align}
One might expect it predicts glass-like segmental localization at high density or low temperature, which has been demonstrated for polymer melts in the context of a more sophisticated theory of collective glassy dynamics \cite{JPCL2013MirigianSchweizer, M2015MirigianSchweizer}. But one also expects it misses mesoscopic (entanglement) localization due to the absence of chain connectivity effects in the force correlations. These expectations are true, as discussed in Appendix II. Thus, for entanglement localization we drop the diagonal contribution of the force memory function matrix since it is dominantly associated with glassy localization which is not of interest. Moreover, for the thread model employed in section IV, the diagonal contribution is of negligible (measure zero) importance.    
\subsection{Center-of-Mass Model}
	 The COM model corresponds to a ``long wavelength'' approximation of a mode-independent memory, $K_p(\infty) \approx K_0(\infty)$. This must \textit{over}predict dynamical constraints which weaken on smaller length scales (higher $p$ index). Adopting this in Eq. (15) yields:
\begin{align}
d_T^2 (\infty) &= \frac{6k_BT}{K_0(\infty)} \frac{1}{N} + \frac{1}{N}\sum_{p=1}^N \frac{12k_BT}{(p\pi/N)^2 + 2 K_0(\infty)} \nonumber \\
&=  \frac{6k_BT}{K_0(\infty)} \frac{1}{N} + \frac{4 \sigma^2}{\pi} \left( \frac{3k_BT}{2K_0(\infty) \sigma^2}\right)^{1/2} \nonumber \\
& \qquad \times  \arctan \left[\pi \left( \frac{3k_BT}{2K_0(\infty) \sigma^2}\right)^{1/2}\right]
\label{eq26}
\end{align}
where the second line follows by approximating the sum as an integral. To proceed requires a closure approximation for $K_0(\infty)$. In the COM model, $\omega_{\alpha \gamma}(k,t) \approx \omega(k) \Gamma_s(k,t)$. This effective $p=0$ model assumes connectivity only affects force correlations via static density correlations, resulting in (as derived in Appendix II): 
\begin{align}
K_{COM}(\infty) &= \frac{\beta^{-1} \rho_s}{3} \int \frac{d\vec{k}}{(2\pi)^3} \left( kC(k) \right)^2 \omega(k) S(k) \nonumber \\
& \qquad \qquad \qquad \times  \exp\left(- \frac{k^2d_T^2}{12} \right) 
\label{eq27}
\end{align}
In Eq.(27), chain connectivity still affects dynamics implicitly via the dependence of force-force correlations on  $\omega(k)$. However, sub-tube scale correlations unique to the $N$ dynamic order parameter matrix theory are lost, i.e., there is no information about off-diagonal dynamic correlations. We will compare below the predictions of the full and COM theories for the tube diameter, which allows us to better understand the role of the full $N$-variable self consistency and the consequences of internal mode dynamic fluctuations on the persistent force correlations that lead to mesoscopic localization.

\section{GAUSSIAN FIELD THEORETIC LIMIT} 
	We now motivate the polymer liquid model adopted to implement our dynamical theory, and present the resultant self-consistent dynamic equations. Entanglement localization emerges on mesoscopic time, $t>\tau_e \sim N_e^2 \tau_0$, and length, $r>r_e \sim \sqrt{N_e} \sigma$, scales  \cite{2003RubinsteinColby,1986DoiEdwards,1979Gennes}. The latter far exceed local structural scales such as $\sigma$, segment hard core diameter ($d$), and the density correlation length, $\xi_p$. But, in reality, entangled dynamics is a consequence of NewtonÕs laws and requires the presence of forces on the local $d$-scale, even if such microscopic information is somehow dynamically blurred. 

\begin{figure}[b!]
\hspace{-0.35in}
 	\includegraphics[height = 2.75 in]{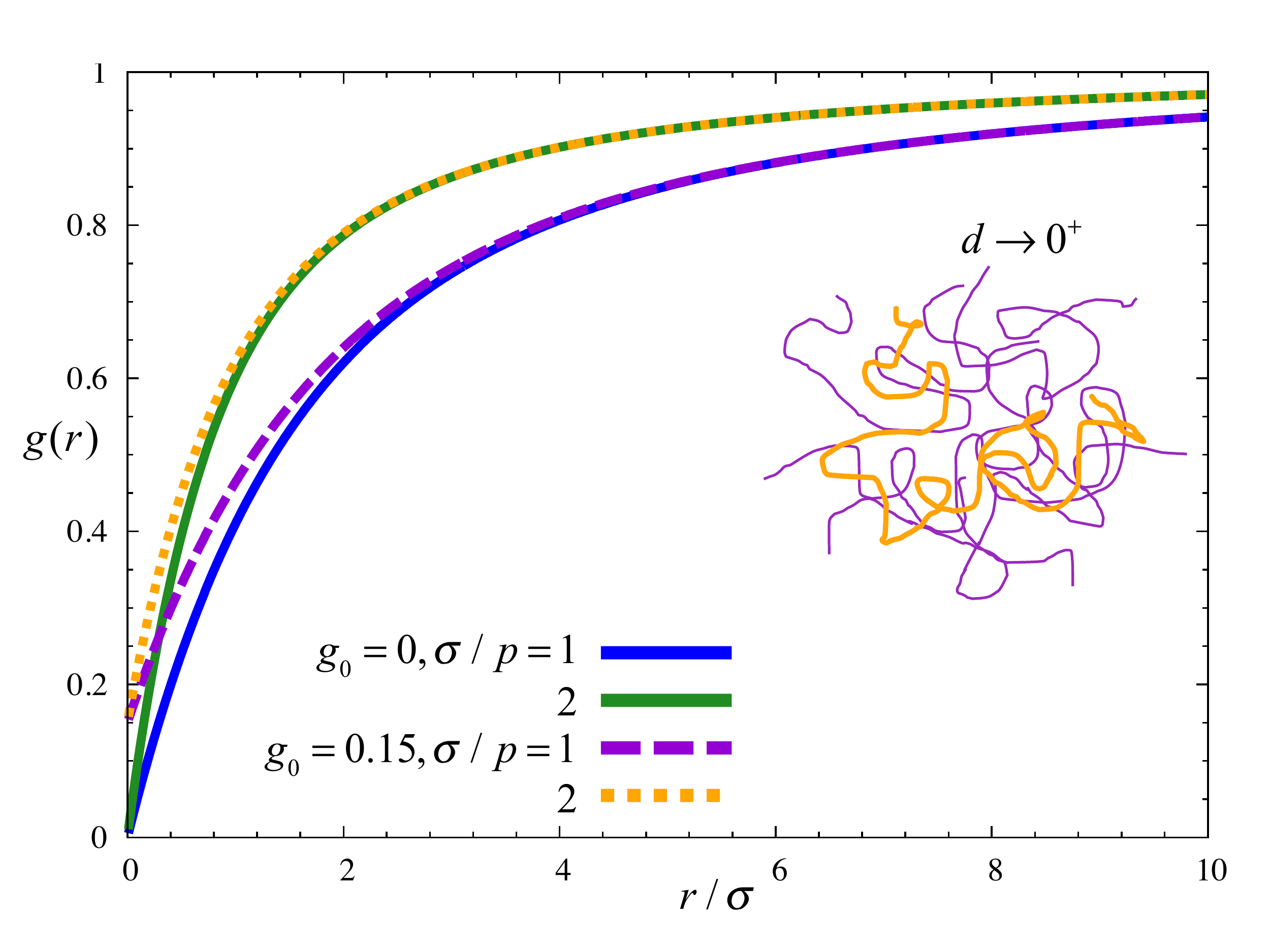}%
	\caption{Interchain site-site (segment-segment) pair correlation function as a function of site separation normalized by segment length for the Gaussian thread PRISM model. Results are shown for $g_0=0$ (solid curves) corresponding to the uncrossable case, and $g_0=0.15$ (dotted and dashed curves) corresponding to a crossable chain, for two values of the dimensionless inverse packing length of $\sigma/p = 1$ and $2$.\label{fig3}}
\end{figure}

How to reconcile the above features is a priori avoided in reptation-tube models. A commonly expressed viewpoint is that structure is irrelevant, and entanglement is a consequence of pure dynamic uncrossability of the trajectories of hypothetical polymer molecules of infinitesimal thickness ($d\rightarrow 0$, ideal gas thermodynamics). This is a non-Hamiltonian description since, at the most fundamental level, interpolymer interactions determine both equilibrium properties and the forces that lead to all dynamics. 

Our first principles approach must consistently treat the equilibrium and dynamical aspects. On the other hand, since entanglement localization is mesoscopic, although $d$ is not literally zero, somehow taking a $d\rightarrow0^+$ limit in the context of coarse graining (not dropping) the local excluded volume constraint should be valid. This motivates our adoption of the so-called universal Gaussian thread model \cite{ACP1997SchweizerCurro, M1988SchweizerCurro} of a single polymer chain and liquid structure. It has been derived from the field theoretic version of the Polymer Reference Interaction Site Model (PRISM) integral equation theory  \cite{ACP1997SchweizerCurro}, as an emergent result of PRISM theory for nonzero thickness chains in semi-dilute solution  \cite{ZPCM1997Fuchs}, and from a Gaussian density fluctuation field theory \cite{PR1993Chandler}. The essential idea is the random walk chain does not intersect other chains based on a point-like ($d\rightarrow0^+$) limit of the no overlap condition (Fig.3 inset).

\subsection{Thread Model and Structural Correlations}

	The PRISM integral equation for the interchain site-site structure of a homopolymer liquid in Fourier space is \cite{ACP1997SchweizerCurro}:
\begin{align}
h(k) &= \omega(k) C(k) \left( \omega(k) + \rho_s h(k) \right) = \omega(k) C(k) S(k)
\label{eq28}
\end{align}
For the Gaussian thread model, the intrachain structure factor is \cite{ACP1997SchweizerCurro, M1988SchweizerCurro}:  
\begin{align}
 \omega(k) &= \frac{1}{N^{-1} + k^2\sigma^2/12}
\label{eq29}
\end{align}
The self contribution (diagonal term) that survives as $k\rightarrow \infty$ is omitted since it is irrelevant for the mesoscopic length scale phenomena of interest. The segment length in Eq. (29) is unambiguously related to polymer chemistry as \cite{ACP1997SchweizerCurro, M1988SchweizerCurro,1969Flory, M1995SchweizerDavid}  $\sigma = \sqrt{C_\infty} l_b$, where $C_\infty$ is the characteristic ratio and $l_b$ is the chemical bond length. 
	
For the direct correlation function, the Percus-Yevick closure \cite{1986HansenMcDonald} is adopted corresponding to its spatial range being the range of the bare interaction. Thus, in the $d\rightarrow0^+$ thread limit, $C(r) = C_0 \delta(\vec{r})$ and thus $C(k) = C_0$. The corresponding collective density fluctuation correlation function is:  
\begin{align}
 S(k) &= \frac{S_0}{1 + k^2\xi_p^2}, \qquad S(r) \propto \frac{e^{-r/\xi_p}}{r\sigma^2}
\label{eq30}
\end{align}
where the dimensionless isothermal compressibility is:
\begin{align}
S_0 &\equiv S(k=0) = \rho_s k_BT \kappa_T \nonumber \\
& = (N^{-1} - \rho_s C_0)^{-1} = 12 (\xi_p/\sigma)^2
\label{eq31}
\end{align}
The structural theory is closed by enforcing a \textit{complete} interchain segment-segment uncrossability constraint:
\begin{align}
 g(r=0) &\equiv g_0 = 0 
\label{eq32}
\end{align}
Simple algebra yields a self-consistent equation for $C_0$, and solving it one has \cite{ACP1997SchweizerCurro}:
\begin{align}
 g(r) &= 1 + \frac{3}{\pi \rho \sigma^2} \frac{\exp[-r/\xi_p] - \exp[-\sqrt{2} r/R_g]}{r} 
\label{eq33}\\
\xi_p^{-1} &= \frac{\pi}{3} \rho_s \sigma^2 + \sqrt{\frac{12}{N\sigma^2}} \approx \frac{\pi}{3p}
\label{eq34}
\end{align}
 where the approximate equality holds for $N>>1$.
	
Characteristic plots of $g(r)$ are shown as the solid curves in Figure 3 for melt-like values of $\sigma/p = 1, \,2$. The dashed and dotted curves show results where the uncrossability condition of Eq.(32) is weakened which we delay discussing until Sec. VI. Significantly, thread PRISM theory correctly  \cite{ACP1997SchweizerCurro, M1988SchweizerCurro, ZPCM1997Fuchs, PR1999FuchsMuller} captures the blob scaling laws in semi-dilute solutions for the density correlation length and osmotic pressure in both good and theta solvents \cite{1979Gennes}: $\xi_p \propto \rho_s^{-\nu}$ and $\Pi \propto \rho_s^{3\nu}$, where $\nu = 1,\, 0.75$, respectively. 

\subsection{Physical Picture and Force Memory Functions}
 	We now elaborate on the physical picture underlying our theory. From Appendix I, the general expression for the force memory function matrix is:
\begin{align}
K_{\alpha \gamma}(t) &= \frac{\beta}{3} \int d\vec{r} \int d\vec{r} \,' \int d\vec{r}\, '' \int d\vec{r}\,'''\vec{F}_\alpha (\vec{r} - \vec{r} \,') \nonumber \\
&\times \omega_{\alpha \gamma}(\vec{r} - \vec{r} \, '', t) \rho_s S(\vec{r}\,' - \vec{r}\, ''', t)\vec{F}_\gamma ( \vec{r} \, '' - \vec{r} \, ''')
\label{eq35}
\end{align}
The cartoon in Fig. 2 indicates the fundamental object is a 4-point in space and time correlation between 2 sites on the tagged polymer and 2 surrounding chains. For the thread model, the effective force is a delta-function, and hence ``force'' enters as a \textit{dynamical contact} between two sites on different chains, as shown in Fig.4. The combination of very short range forces and density correlation length implies that at $t=0$ that all 4 sites must be close in space (Fig. 4a). This means the sites $\alpha$ and $\gamma$ on the tagged chain are spatially close, in a configuration that is essentially a self-intersection which will constraint the $\left| \alpha - \gamma \right| -2$ connected sites (loop) between them. \textit{If} kinetic arrest occurs on a length scale $d_T$ at long times, then the tagged and matrix chains will displace over a mesoscopic distance $\sim d_T$.  Hence, for force correlations, such a relaxation and localization on the tube diameter scale determines the amplitude of persistent long time spatial correlations (Fig.4b). The physical picture underlying the force memory function matrix is thus of arrested (long lived in practice) coarse-grained ($d_T$ scale) dynamic contacts between segments on a pair of interpenetrating chains. This picture seems qualitatively consistent with the idea of persistent dynamic contacts deduced from simulation studies \cite{PR1996Ben-NaimGrest, *JCP1997SzamelWang, SM2014Likhtman, *M2014LikhtmanPonmurugan, PR2012Everaers, *M2014QinMilner, *M2012AnogiannakisTzoumanekas}. An alternative interpretation is discussed in Appendix III which buttresses the above discussion.  

\begin{figure}[t!]
 	\includegraphics[height = 1.9 in]{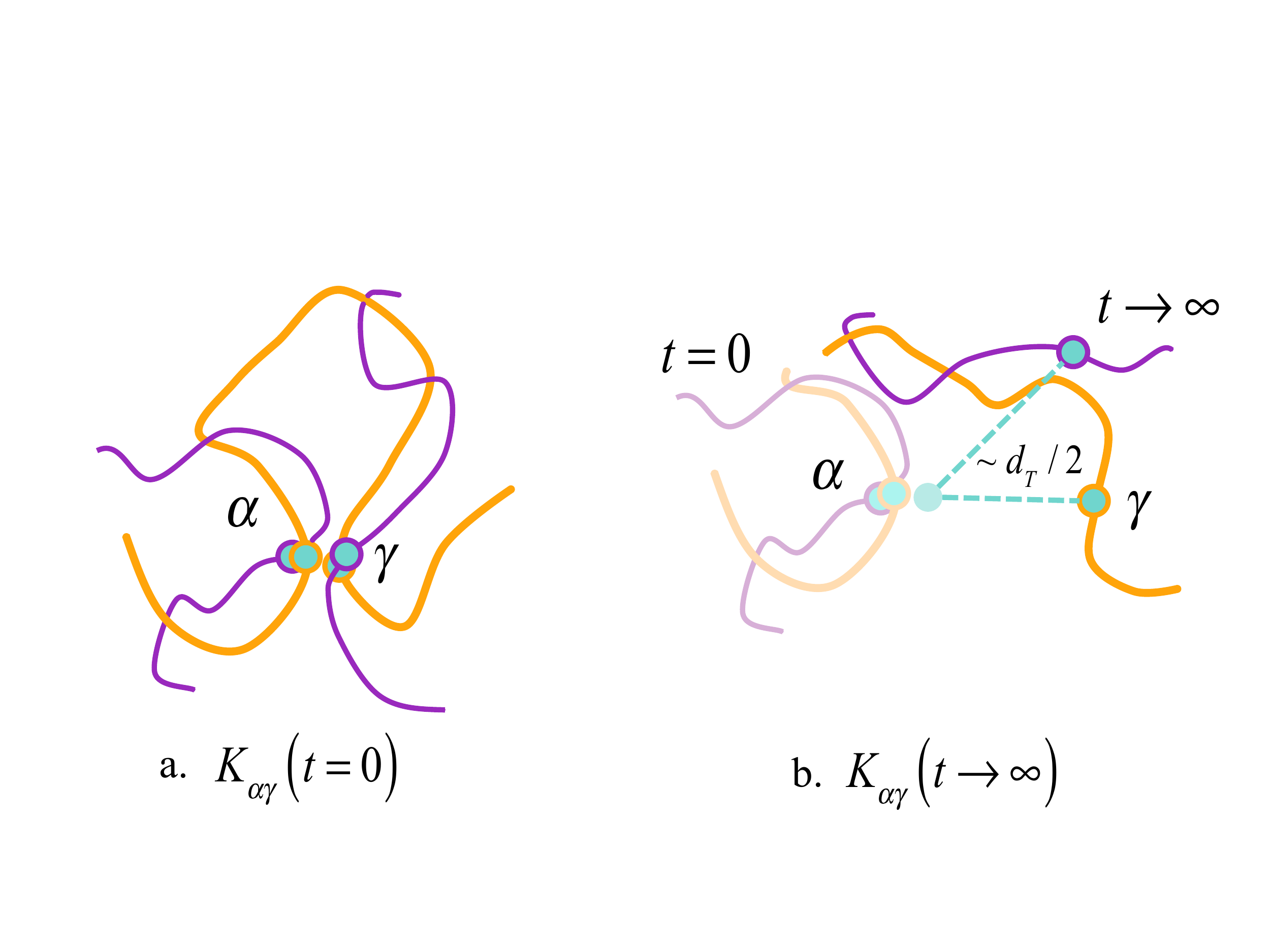}%
	\caption{Schematic of the dynamic force-force correlations at short and long times based on the Gaussian thread polymer model of structure. a) Due to the point-like nature of interchain interactions, at time $t=0$ the force correlations must involve four segments very close in space. Two are on the tagged (orange) chain, $\alpha$ and $\gamma$, and the other two (in literal contact with the tagged chain) are on matrix (purple) chain(s). (ii) In the kinetically arrested (entangled) state at long times, segments relax only on a scale determined by the mesoscopic dynamic localization length or tube diameter, $d_T$.\label{fig4}}
\end{figure}

The equations that define the dynamical theory simplify in the thread limit:
\begin{align}
K_{\alpha \gamma}(t) &= \left(1-\delta_{\alpha\gamma}\right) \frac{\beta^{-1}}{3} \rho_s C_0^2 S_0 \nonumber \\
 & \qquad \times \int \frac{d\vec{k}}{(2\pi)^3} k^2 \omega_{\alpha \gamma}(k,t) \frac{\Gamma_c(k,t)}{1+k^2 \xi_p^2}
\label{eq36}
\end{align}
In the simpler COM model, an explicit expression can be derived (see Appendix III): 
\begin{align}
K_{COM}(\infty) &= \frac{12 \sqrt{3\pi} k_BT}{\pi^2 p d_T^3 \sigma^4} C_0^2 S_0 \left[ 1 - \frac{d_T^2}{3R_g^2}  \right . \nonumber \\
& \left .+ \frac{1}{3} \sqrt{\frac{\pi}{6}} \frac{d_T^3}{R_g^3} e^{d_T^2/6R_g^2} \left( 1 - \mbox{erf}\left[\frac{d_T}{\sqrt{6} R_g} \right]\right) \right]
\label{eq37}
\end{align}
The numerical results presented in the next section for the COM model can be analytically understood, which provides additional insight to why and how our prediction of mesoscopic localization first emerges and its consequences for the tube diameter. 

\section{RESULTS: UNCROSSABLE CONNECTED CHAINS}
	Based on the structural model of Section IV, the parameter inputs to the theory are only the chain length $N$ (or $R_g$), and the segmental volume fraction, \linebreak $\eta_\sigma \equiv \rho_s \pi \sigma^3/6 = \pi \sigma/6p$, which depends on polymer concentration and chemistry. Our primary focus is melts; semi-dilute solutions are briefly analyzed in the final sub-section.  

\subsection{Dynamic Localization Transition and Length Scale}
To model polymer melts we choose two packing lengths which essentially span the entire range for synthetic chain polymers: $\sigma/p = 1 \, (\eta_\sigma = 0.52)$ and $\sigma/p = 2$ $(\eta_\sigma = 1.05)$. These choices are (using $\sigma = \sqrt{C_\infty} \, l_b$) representative of polystyrene and polyethylene, respectively, which have packing lengths $p\sim 0.4$ and 0.17 nm  \cite{M1994FettersLohse}  

\begin{figure}[t!]
\hspace{-0.35in}
 	\includegraphics[height = 2.75 in]{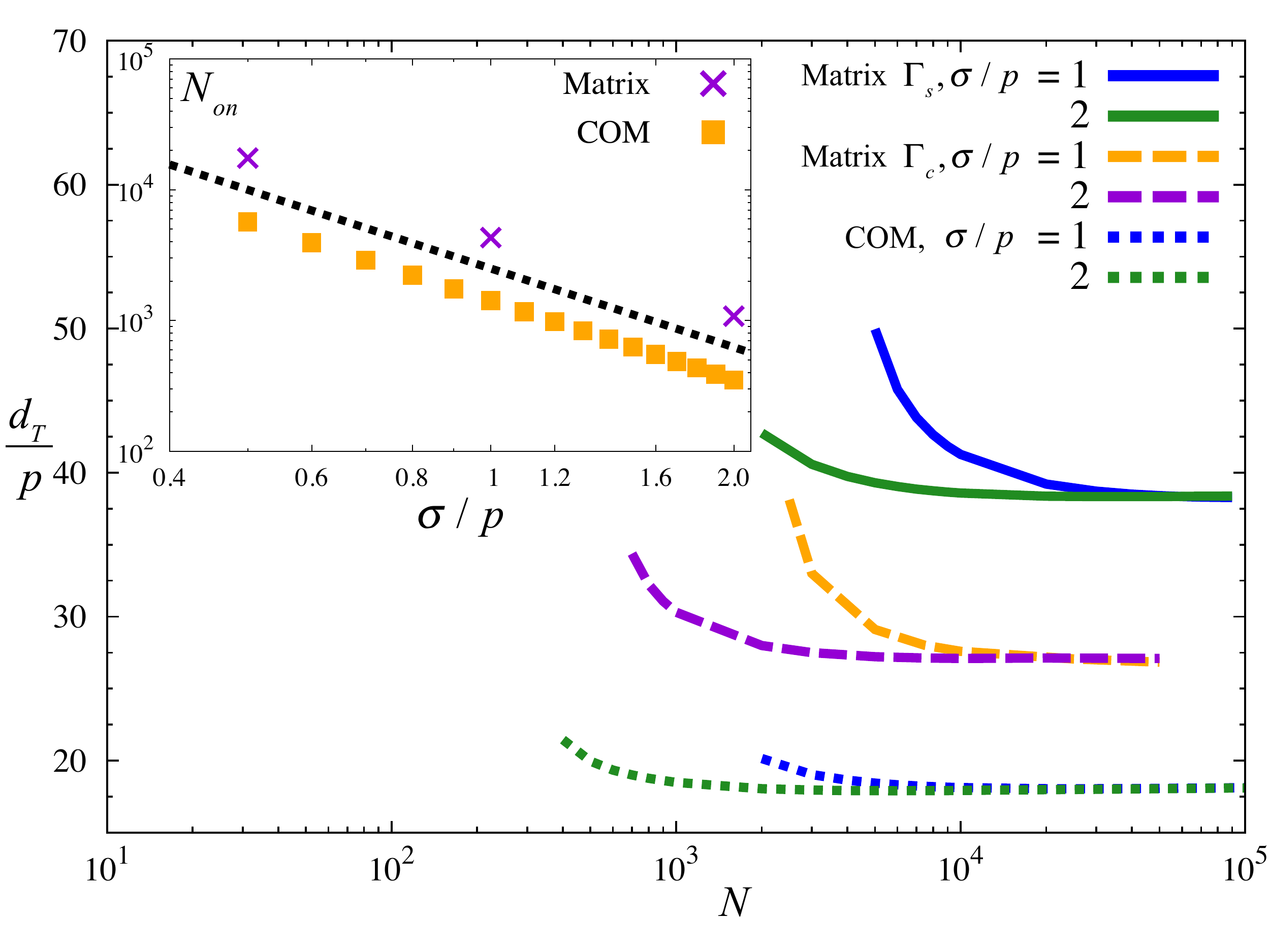}%
	\caption{(main) Tube diameter, normalized by the packing length, as a function of chain length. The solid curves show the full matrix dynamical theory results using the segmental Vineyard closure (Eq. (23)), the dashed curves show the analogous results using the chain Vineyard closure (Eq. (24)), and the dotted curves show the simpler COM theory results. In all three cases, results are presented for $\sigma/p = 1$ (blue and orange, right curves) and 2 (green and purple, left curves). (inset) The mesoscopic localization onset chain length as a function of the dimensionless inverse packing length for the dynamical matrix theory with segmental Vineyard closure (crosses) and the COM theory (squares). The dotted line shows the power law $N_{on} \propto (\sigma/p)^{-2}$.\label{fig5}}
\end{figure}

	The main frame of Figure 5 shows results for the tube diameter normalized by the packing length as a function of $N$. The solid curves show the results of the full matrix calculation using the segmental Vineyard closure (Eq. (23)), the dashed curves show the results of the matrix theory with chain Vineyard closure (Eq. (24)), and the dotted curves show the COM model results. In all cases, for small enough $N$, \textit{no} mesoscopic localization is predicted, per the Rouse model of unentangled liquids. At a critical chain length $N_{on}$, there is a discontinuous localization transition, akin to the classic concept \cite{2003RubinsteinColby, 1986DoiEdwards,1979Gennes, PPS1967Edwards, JCP1971Gennes, AP2002McLeish} of the existence of a well-defined value of $N_e$. Significantly, the emergence of mesoscopic localization always occurs on the chain size scale, $d_{T, on} \approx (1-2) R_g$. The tube diameter then modestly decreases with $N$ until it reaches a limiting value where $d_T \propto p$. Quantitatively, for the segmental Vineyard full matrix theory  $d_T = 38 p$, for the chain Vineyard matrix theory $d_T = 27 p$, and for the COM model $d_T \approx 18 p$. The fact that the COM model predicts the smallest $d_T$ and $N_on$ is expected since it ignores weakening of the confining force correlations with decreasing length scale. The COM result is analytically derivable (see Appendix III):
\begin{align}
d_T = 18 \sqrt{\frac{3}{\pi}} \, p \approx 17.6 p
\label{eq38}
\end{align}
From the analysis in Appendix III one can understand the self-consistent competition that leads to abrupt mesoscopic localization at a critical value of $N$ as a consequence of a growing number of dynamically constrained internal conformational modes. 

The above results agree with the experimental finding \cite{M1994FettersLohse} of $d_T \approx 18 p$ to within a factor of 2 or better. Phenomenological arguments have been advanced  \cite{PRL1987KavassalisNoolandi, *M1987Lin, PY1989WittenMilner}  that motivate the proportionality $d_T \propto p$, but they provide no quantitative insight to the mesoscopic size of the dynamic length scale. In contrast, the mesoscopic nature of the tube diameter is a \textit{bone fide} prediction of our approach, which also provides a fundamental basis for the packing model idea \cite{M1994FettersLohse, S2004EveraersSukumaran, PRL1987KavassalisNoolandi, *M1987Lin, PY1989WittenMilner}. The near exact agreement of the COM model with experiment is accidental. The entanglement onset occurs at $N_{on} = 6R_g^2/\sigma^2 \approx 6 d_{T, on}^2 / \sigma^2 \approx 6 N_e$ . This is a nontrivial result in the sense that dynamic localization cannot occur unless the chain is on average larger than the tube diameter, or equivalently $N_{on} > N_e$. The precise relationship between $N_{on}$ and the packing length is shown in the inset of Fig. 5 for melt-like and semi-dilute packing fractions. The crosses (squares) show the full matrix (COM) theory results. In both cases $N_{on} \propto (\sigma/p)^{-2} \propto (\sigma/d_T(\infty))^2$ (dotted black line), per existing phenomenology \cite{2003RubinsteinColby, 1986DoiEdwards,1979Gennes}. 

\subsection{Spatial Structure of Arrested Conformation and Force Correlations}
	The $N$ order parameter matrix theory predicts off-diagonal dynamic correlations which provide deeper insight into the kinetically-arrested structure of a localized polymer. For these properties, we find that the segmental and chain Vineyard approximations give virtually identical results when normalized in the manner presented below. Thus, only results based on the simpler segmental Vineyard closure are shown. 

Calculations of the chain dynamic second moment matrix of Eq. (15) are shown in Figure 6. All results roughly collapse when the real space segmental separation is scaled by the tube diameter, $\left | \alpha - \gamma \right | \sigma^2/d_T^2$. As the segment separation grows, the dynamic correlations decrease roughly exponentially on the scale of  the tube diameter:
\begin{align}
\delta \mu_{\alpha \gamma}^{(2)} (\infty) &= \frac{d_T^2}{4} \exp \left( -\left | \alpha - \gamma \right | \sigma^2/0.4 \,d_T^2\right)
\label{eq39}
\end{align}
Eq. (39) quantifies the strong dynamic correlations on scales inside the mean localization length. Beyond the tube scale, the correlations are very small ($< 0.01$). 
\begin{figure}[b!]
\hspace{-0.35in}
 	\includegraphics[height = 2.75 in]{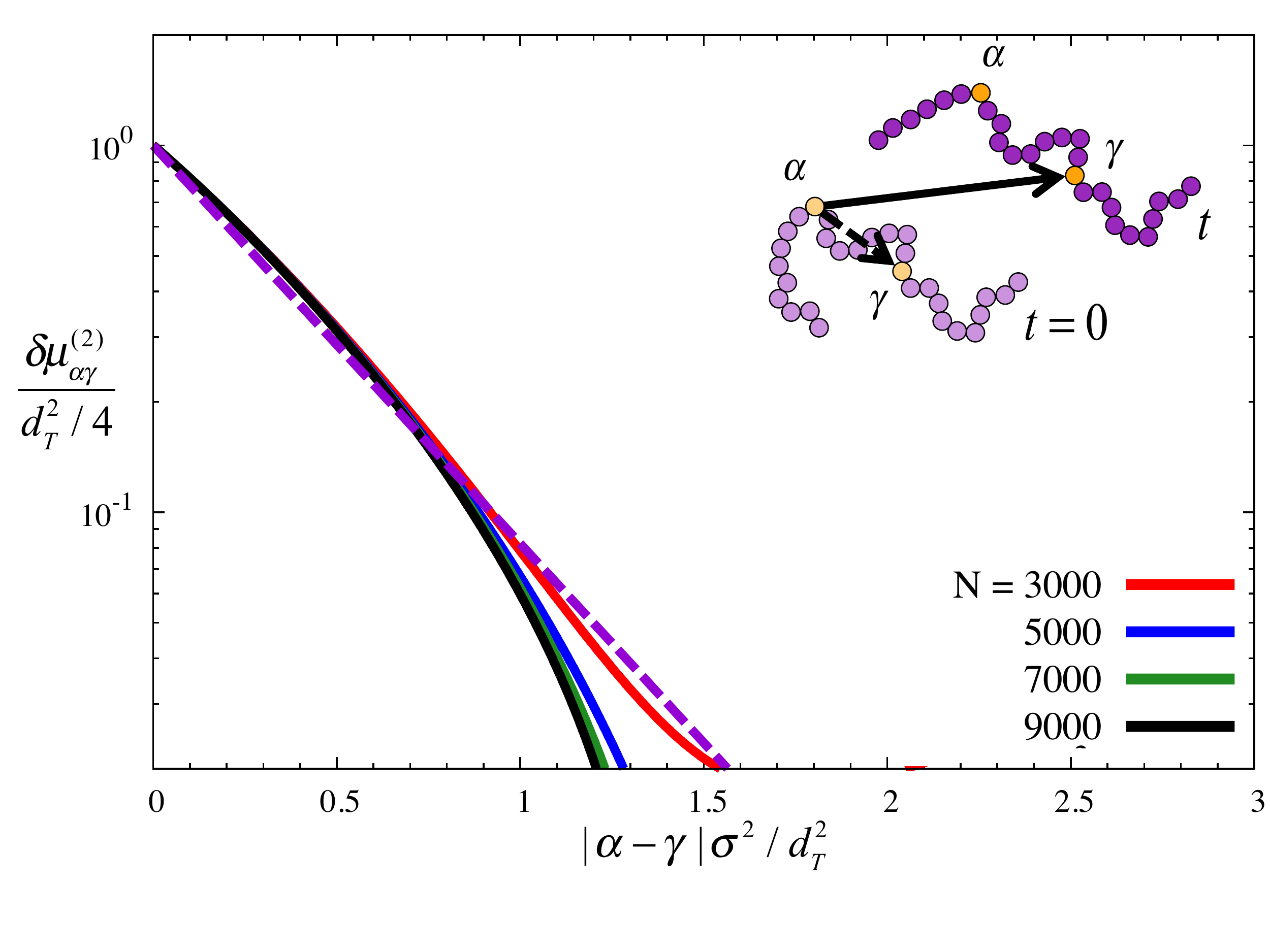}%
	\caption{The dynamic portion of the chain second moment, normalized by its zero separation value of $\delta \mu_{\alpha \alpha}^{(2)}=d_T^2/4$, is plotted as a function of the segment separation normalized by the tube diameter squared, for a fixed value of $\sigma/p=2$. The solid curves show the theoretical results for the indicated chain lengths, while the dashed curve shows the exponential form of Eq.(39).\label{fig6}}
\end{figure}
	
 Figure 7 shows the dynamically-arrested force-force correlations in segment (inset) and Rouse mode (main frame) space. The Rouse mode space $K_p$, normalized by its COM ($p=0$) value, is shown as a function of the mode index normalized by the degree of entanglement, $N_e/N = d_T^2 /N \sigma^2$. These arrested force correlations define via the GLEs an effective spring constant on a scale $\Lambda_p \sim \sqrt{N/p} \, \sigma$. The results for all chain lengths roughly collapse. With decreasing length scale (increasing mode index), Fig.7 shows the force correlations decrease exponentially over roughly the first order of magnitude of decay:
\begin{align}
K_p \approx K_0 \exp(-pd_T^2/N \sigma^2)
\label{eq40}
\end{align}
This decay is modest over the wide range of mode index values of $p=0\rightarrow N/N_e$.  	

The analogous real space force correlations are shown in the inset of Fig. 7 as a function of normalized segment separation. All curves again collapse. These correlations can be divided into two regimes. The first is a short range exponential decay:
\begin{align}
K_{\alpha \gamma}(\infty) &= K_{\alpha \alpha} (\infty) \exp ( -\left| \alpha - \gamma \right| \sigma^2 / 0.44 d_T^2) \nonumber \\
K_{\alpha \alpha}(\infty) &= \sqrt{\frac{\pi}{6}} \frac{k_BT \sigma^4}{8 pd_T^5} \propto \frac{k_BT}{\sigma^2} \left(\frac{\sigma}{p}\right)^6
\label{eq41}
\end{align}
After a decade and a half of decay, there is a crossover to a power law behavior:
\begin{align}
K_{\alpha \gamma}(\infty) &= K_{\alpha \alpha} (\infty) \left[ \frac{\left| \alpha - \gamma \right| \sigma^2}{ d_T^2}\right]^{-5/2} , \;\; \left| \alpha - \gamma \right| \sigma^2 >> d_T^2
\label{eq42}
\end{align}
This result can be analytically derived (Appendix III). The inverse power law decay is reminiscent of ``long time tails'' that emerge in various physical systems for diverse reasons \cite{1986HansenMcDonald,2001Zwanzig}. Here it is due to the fractal (long range) nature of the single chain connectivity constraints on the dynamically arrested spatial polymer density distribution. 

\begin{figure}[t!]
\hspace{0.0 in}
 	\includegraphics[height = 2.65 in]{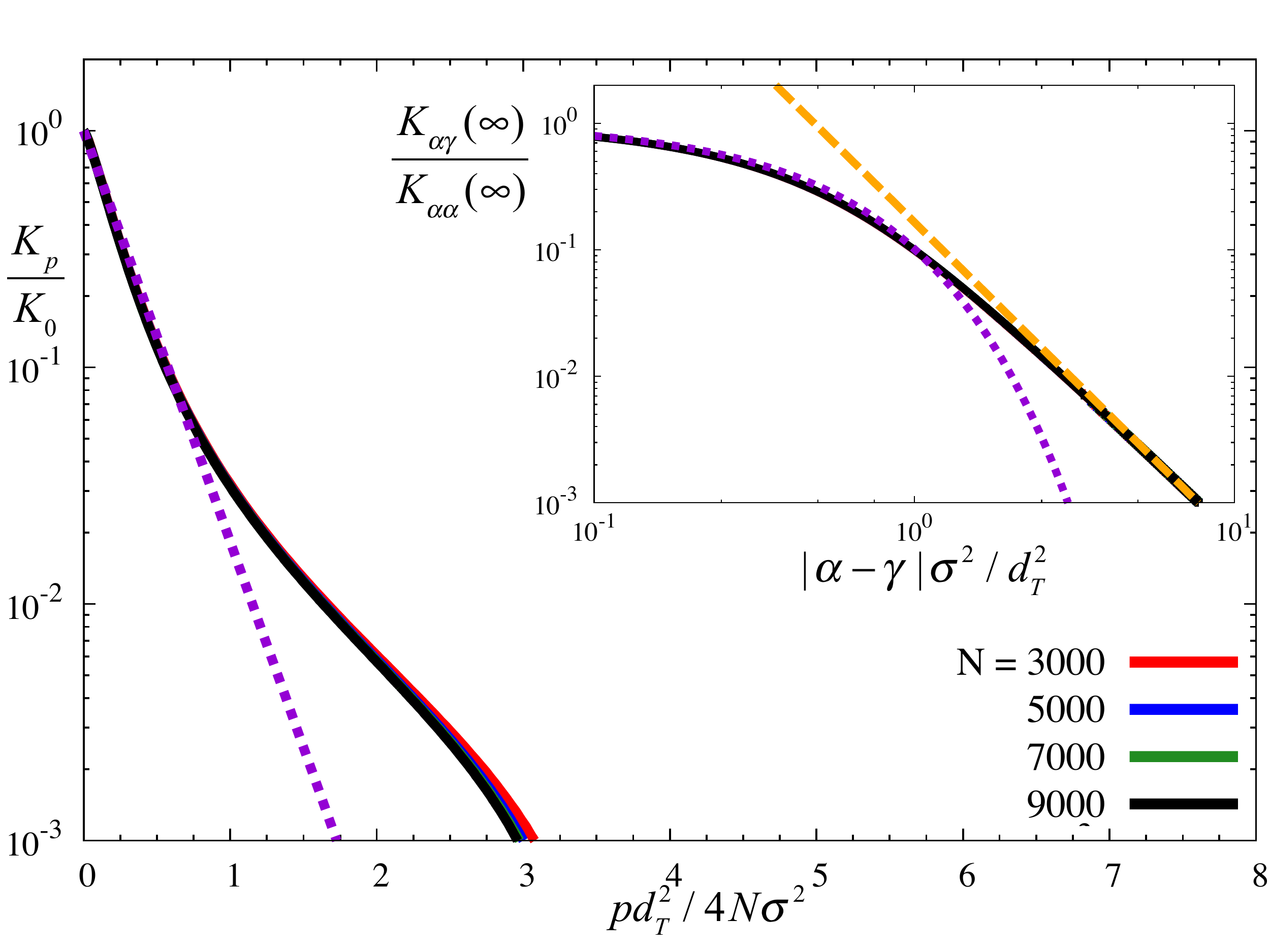}%
	\caption{(main) The kinetically arrested values of the Rouse mode force correlation function normalized by the center-of-mass ($p=0$) result, as a function of mode index normalized by degree of entanglement, $N/N_e \equiv 6R_g^2/d_T^2$, at fixed $\sigma/p=2$. The solid curves for the indicated chain lengths all collapse; the dotted curve shows the exponential of Eq.(40). (inset) Same as main frame but in real space as a function of the segment separation normalized by the tube diameter. Good collapse is again found for various chain lengths. The dotted purple curve shows the exponential of Eq.(41), while the dashed orange curve shows the power law tail of Eq.(42). \label{fig7}}
\end{figure}

\subsection{Dynamically Arrested Coherent Single Chain Correlations}
The arrested coherent dynamic structure factor of a tagged chain is the $t \rightarrow \infty$ limit of Eq.(19). This quantity is measurable in neutron spin echo experiments \cite{AP2002McLeish, NSEPS2005RichterMonkenbusch}, or via simulations where the ergodicity-restoring reptation motion can be turned off by hand. The main frame of Figure 8 shows results in a doubly normalized representation for $\sigma/p = 2$ and indicated chain lengths. All curves collapse. The dotted yellow curve shows a Gaussian model based on the segmental dynamic density correlations $\Gamma_s(k, \infty)$ of Eq. (23), which is the main contribution. This is the primary reason that the segmental and chain Vineyard approximations for collective dynamics yield very similar results.  
	
\begin{figure}[t!]
\hspace{-0.35 in}
 	\includegraphics[height = 2.75 in]{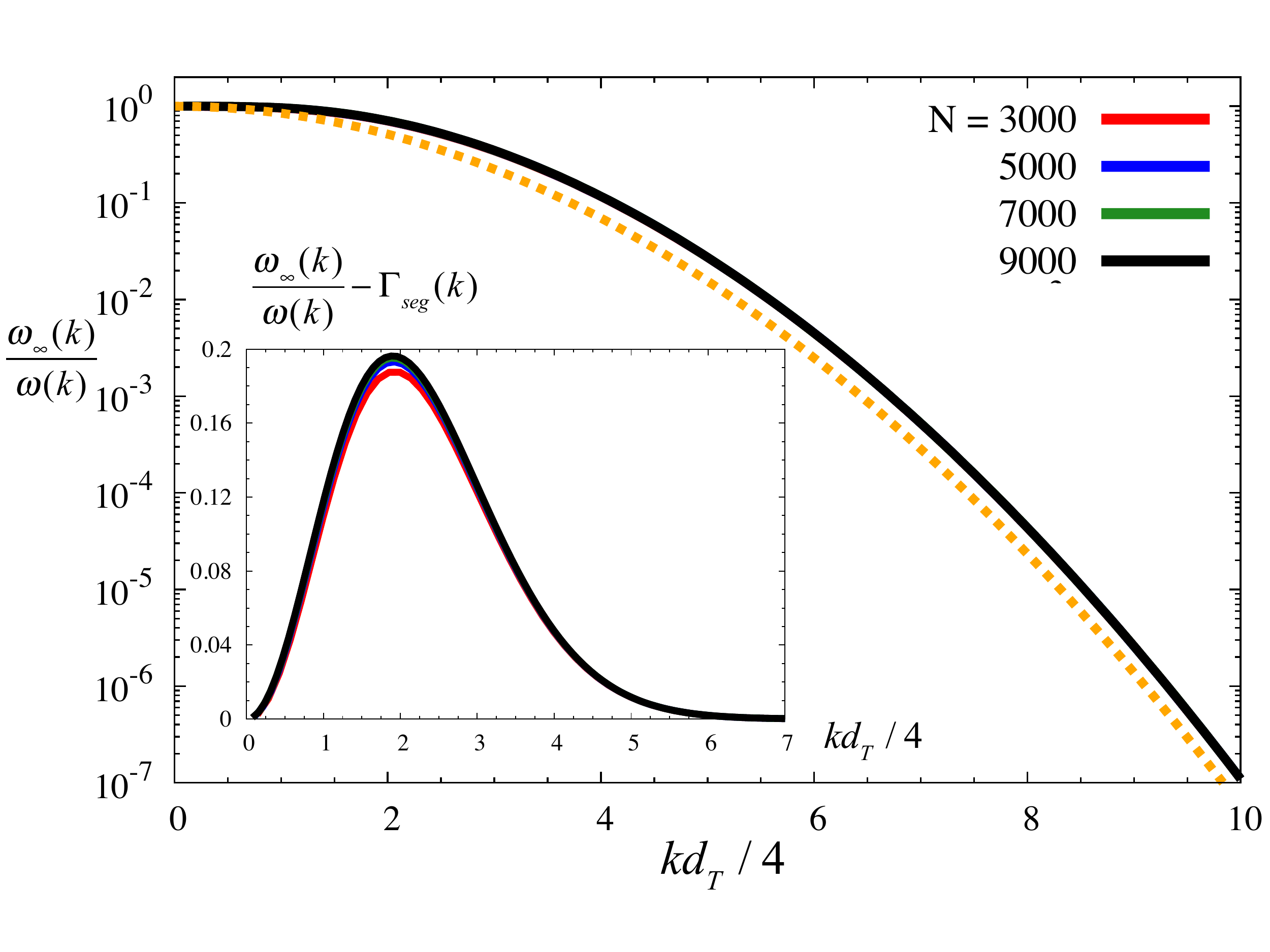}%
	\caption{(main) The kinetically arrested collective single chain dynamic structure factor normalized by its equilibrium value as a function of non-dimensionalized wave vector for $\sigma/p=2$. The solid curves show results for various chain lengths and all roughly collapse. The dotted yellow curve shows the segmental Gaussian $\Gamma_s(k)$ of Eq.(23). (inset) Same results after subtracting the segmental contribution. \label{fig8}}
\end{figure}

We have also calculated the arrested bond-bond correlations defined by:
\begin{align}
\left< \vec{r}_\alpha (\infty) \cdot \vec{r}_\gamma (\infty)\right> & \equiv \left< \left(\vec{R}_\alpha (\infty) - \vec{R}_{\alpha -1}(\infty)  \right) \right . \nonumber \\
&\qquad \cdot \left . \left(\vec{R}_\gamma (\infty) - \vec{R}_{\gamma -1}(\infty) \right)\right> \nonumber \\
& = \sqrt{\frac{2}{N}} \, \sum_{p=1}^N \left< \vec{x}_p^2 (\infty)\right> \psi_p (\alpha - \gamma) \nonumber \\
& \qquad \times \left[ 1 - \cos \left( \frac{p\pi}{N}\right)\right]
\label{eq43}
\end{align}
Solving Eq.(13) (Appendix I), Eq. (43) is evaluated. We find (not shown) the bond correlations are nearly diagonal $\left< \vec{r}_\alpha (\infty) \cdot \vec{r}_\gamma (\infty)\right> \approx \sigma^2 \delta_{\alpha \gamma}$, and hence randomly orientated as in equilibrium. Such a simple result is likely partially, or largely, a consequence of the Gaussian factorization of multi-point dynamic correlations.

\subsection{Plateau Shear Modulus}
	The entanglement shear modulus can be ÒrigorouslyÓ computed from the Rouse model expression for the single chain entropic stress relaxation modulus  \cite{1986DoiEdwards}:
\begin{align}
G_{Rouse} &= \frac{\rho_p}{k_BT} \sum_{\alpha, \gamma =1}^N \left< F_\alpha^x(\infty) R_\alpha^y(\infty) F_\gamma^x(0) R_\gamma^y(0)\right> \nonumber \\
&= \rho_s k_BT \frac{1}{N} \sum_{p=1}^N \frac{\left< \vec{x}_p(\infty) \cdot \vec{x}_p(0)\right>^2}{\left<x_p^2\right>^2}
\label{eq44}
\end{align}
where $F_\alpha^i$ and $R_\alpha^i$ are the net force on and position of segment $\alpha$ in the $i$th direction, respectively, and the numerator on the second line is the arrested mode amplitude of Eqs. (13). A second intuitive model for the plateau modulus adopts the crosslinked rubber picture often employed to empirically estimate $N_e$  \cite{2003RubinsteinColby, 1986DoiEdwards, 1979Gennes}:
\begin{align}
G_{Ent} &= \frac{\rho_sk_BT}{N_e} = \frac{\rho_s k_BT \sigma^2}{d_T^2}
\label{eq45}
\end{align}
\begin{figure}[b!]
\hspace{-0.0 in}
 	\includegraphics[height = 2.65 in]{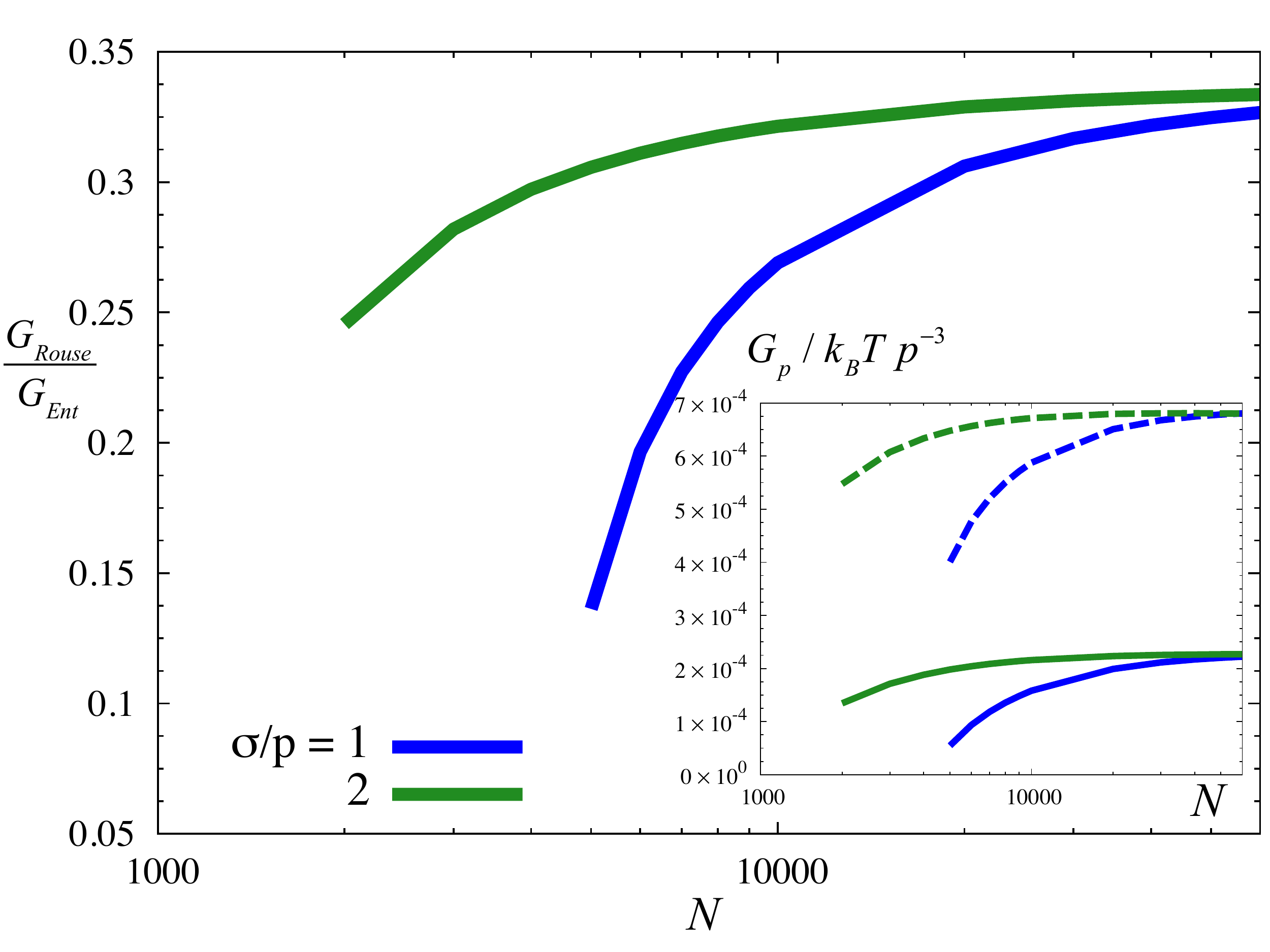}%
	\caption{(main) Ratio of the entanglement plateau shear modulus of Eq.(44) divided by its rubbery model analog of Eq.(45). Results are shown as a function of chain length $N$ for two melt-like dimensionless inverse packing lengths. (inset) The two moduli expressed in units of $k_BTp^{-3}$. The solid curves show $G_{Rouse}$ which plateaus at $2.3\times 10^{-4}$, the dotted curves show $G_{Ent}$ which plateaus at $6.8\times10^{-4}$.  \label{fig9}}
\end{figure}

Figure 9 shows typical results. The inset normalizes the modulus by $k_BT/p^3$, where the solid (dashed) curves show $G_{Rouse}$($G_{Ent}$). In all cases the modulus is roughly independent of chain length. Equation (45) and the theoretical result $d_T = 38 p$ yield:
\begin{align}
G_{Ent} &= (6.8\times 10^{-4}) \frac{k_BT}{p^3}
\label{eq46}
\end{align}
 This is consistent with experiments and simulations (Eq.(1)) which find $G_{expt} = (2.3\times 10^{-3}) k_BT/p^3$ \cite{M1994FettersLohse, S2004EveraersSukumaran}. The quantitative deviations are due to the modest numerical discrepancy in the predicted tube diameter. The plateau moduli computed in the two ways are roughly proportional (Fig. 9 main). Near onset (crossover regime), they have, unsurprisingly, slightly different chain length dependences. However, far from the onset, in the large $N$ limit, $G_{Ent} \approx 3G_{Rouse}$.  

\begin{figure}[b!]
\hspace{-0.35 in}
 	\includegraphics[height = 2.75 in]{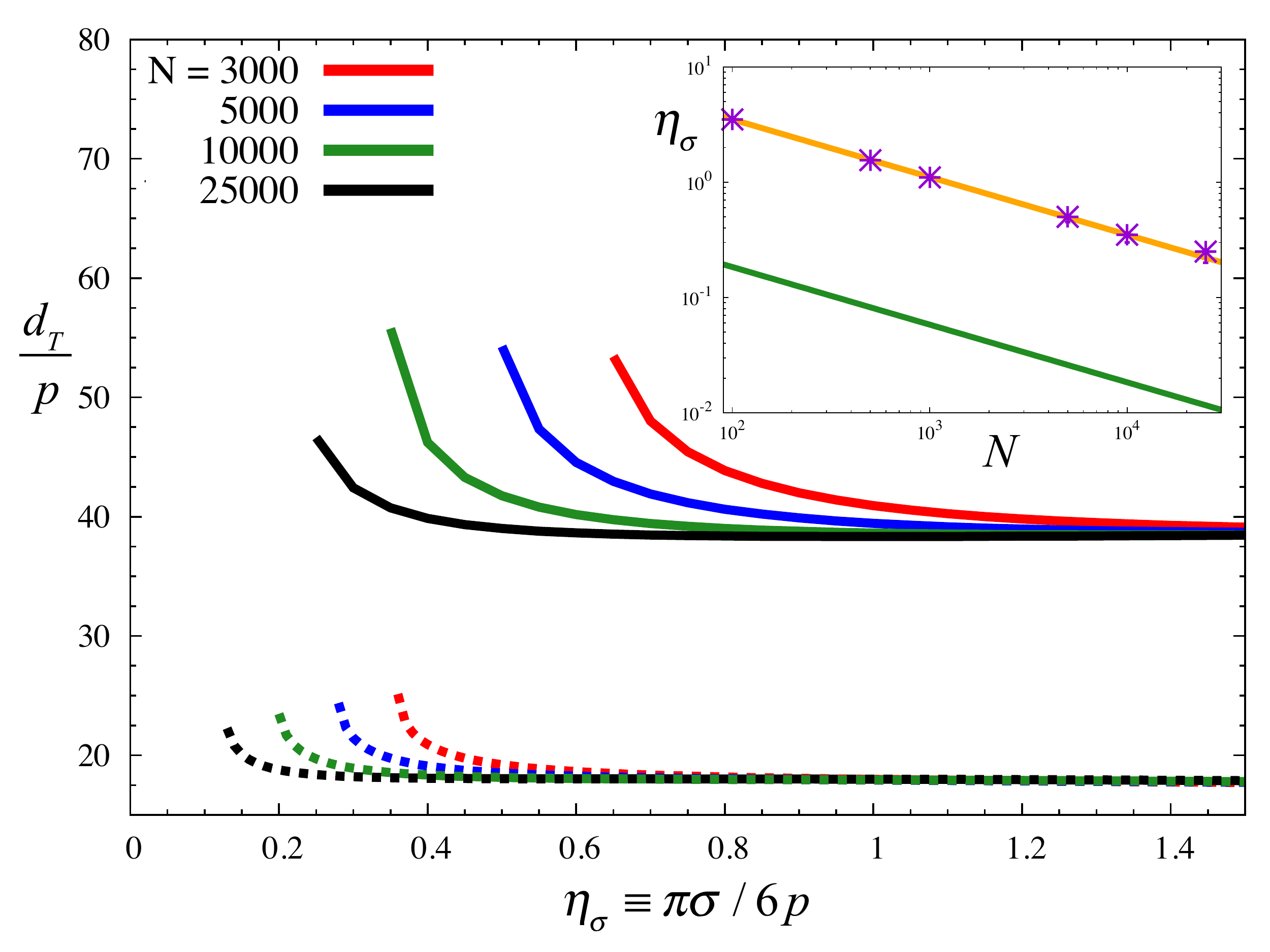}%
	\caption{(main) Tube diameter normalized by packing length in semi-dilute solutions and melts as a function of segmental volume fraction.  The solid curves show results for the full matrix dynamical calculations with the segmental Vineyard closure, while the dotted curves show the corresponding COM theory result. Results for 4 chain lengths (increasing from right to left) are shown. (inset) Polymer volume fraction at the onset of entanglement localization (points) as a function of chain length for the segmental Vineyard dynamic matrix theory. The green (lower) line is the semi-dilute crossover $\eta_{\sigma}^*$, the orange line through the data is $19\eta_{\sigma}^*$.\label{fig10}}
\end{figure}

\subsection{Semi-Dilute Solutions}
We now study semi-dilute solutions within the thread model description. Since chains repel, strictly speaking it applies to good (not theta) solvents since the second virial coefficient is nonzero. It should be viewed as an effective Gaussian model\cite{ACP1997SchweizerCurro}, as employed in Edwards-like field theories \cite{1986DoiEdwards, PPS1966Edwards}, where deviations from ideal conformation is mimicked by allowing $\sigma \propto N^{0.1}$ in dilute solution and $\sigma \propto \rho_s^{-1/8}$ in semi-dilute solution  \cite{2003RubinsteinColby, 1986DoiEdwards, 1979Gennes}. Our focus is the polymer density dependence at fixed chain length. 

The main frame of Figure 10 plots the normalized tube diameter as a function of volume fraction $\eta_\sigma \equiv \rho_s \pi \sigma^3 = \pi \sigma/6p$. The solid (dotted) curves show results from the full dynamical matrix theory with the segmental Vineyard closure (COM). At low enough volume fraction, no mesoscopic localization occurs. At a critical volume fraction, localization emerges and abruptly recovers the melt result, $d_T = 38 p$  $(18 p)$ for the matrix (COM) theory, with increasing volume fraction. 

Physically, entanglement localization must emerge only at concentrations beyond the the dilute to semi-dilute crossover at (for $\sigma \propto \rho_s^0$) $\eta_\sigma^* = (3\sqrt{6}/4) N^{-1/2}$, which is shown by the green (lower) line in the inset of Figure 10. For $\eta_\sigma \gtrsim \eta_\sigma^*$, the onset volume fractions predictions are indicated by the stars in the inset of Fig. 10, which is a re-plotting of the results in the inset of Fig. 5. The inset of Fig. 10 shows the theoretical results for several chain lengths. For all chain lengths studied, $\eta_{on} = 19 \eta_{\sigma}^*$, per the solid yellow line in Fig. 10; within the effective Gaussian chain model, the same result applies to good solvents. Thus, our result agrees well with experiments  \cite{2003RubinsteinColby} which find $\eta_{on} \approx (4-30) \eta_\sigma^*$. An analytic derivation of this result is given in Appendix III. 

\section{CONSEQUENCES OF DYNAMIC CHAIN CROSSABILITY}
We now examine more deeply the main physical thesis of our approach -- the dynamical phenomenon called entanglement that leads to mesoscopic localization and rubber-like elasticity can be captured if one takes into account local chain uncrossability (force contacts), long range chain connectivity, and self-consistency between conformational dynamics and interchain dynamic force correlations over \textit{all} length scales. As relevant background and motivation, we discuss three key points.

First, a common criticism of \textit{all} microscopic attempts to describe entanglement effects is they do not enforce dynamic uncrossability at \textit{all} times. Of course, this impossible for any tractable theory, even for the ``cage effect'' associated with glassy dynamics in liquids of spherical particles. We believe that the germane question is whether the enforcement of such exact uncrossability is necessary to capture the essence of entanglement localization. The answer is not obvious; indeed, self-consistent pair-level theories (e.g., mode-coupling  \cite{2008Gotze}) that do not satisfy point (i) can still capture statistical caging and emergent localization associated with glass and gel physics.  

Second, although obvious, in the tube model the concept of dynamic ``uncrossability'' is a highly coarse grained, soft notion where chains cross at all times before ``entanglements emerge'' and on all scales less than the tube diameter. Such crossing is unphysical, but nonetheless the phenomenological tube model can successfully capture coarse-grained \textit{consequences} of dynamic uncrossability \textit{without} strictly enforcing it on the microscopic length scale it exists for real polymer molecules. 

Third, some simulation-based attempts aimed at better understanding what an entanglement is have concluded the key is very slowly relaxing or statistically persistent ``contacts'' between a pair of long intertwined chains in a dense liquid  \cite{PR1996Ben-NaimGrest, *JCP1997SzamelWang, SM2014Likhtman, *M2014LikhtmanPonmurugan, PR2012Everaers, *M2014QinMilner, *M2012AnogiannakisTzoumanekas}. Such persistent contacts are the centerpiece of our dynamical theory, per Figure 4.

\subsection{Simulation Studies}
The chain crossability issue has been studied with molecular dynamics (MD) simulations based on forces and NewtonÕs law  \cite{PSCR2012Likhtman, PRL1991DueringKremer, M2014KalathiKumar, PNAS2015ZhangWolynes}, and Monte Carlo simulations \cite{JCP1994Shaffer, JSP2011WittmerCavallo} based on dynamical moves. Our theory works at the force level, so MD studies are most relevant. They often employ bead-spring Kremer-Grest model \cite{PRL1991DueringKremer} where segments interact pairwise via the repulsive Weeks-Chandler-Anderson \cite{JCP1971WeeksChandler} potential:
\begin{align}
U_{WCA}(r) &= \left\{ 
\begin{array}{cc}
4\epsilon \left[(\sigma/r)^{12} - (\sigma/r)^6 + 1/4\right] & r\leq2^{1/6} \sigma \\
0 & r>2^{1/6} \sigma
\end{array}
\right.
\label{eq47}
\end{align}
 where the $\epsilon$ is the energy scale. Nearest neighbors are connected via a nonlinear spring:
\begin{align}
U_{WCA}(r) &= \left\{ 
\begin{array}{cc}
-0.5 R_0^2 k \ln  \left[1-(r/R_0)^2\right] & r<R_0 \\
\infty & r\geq R_0
\end{array}
\right.
\label{eq48}
\end{align}

 	By choosing $R_0 = 1.5\sigma$ and $k = 30 \epsilon/\sigma^2$, it was found that chains do not cross and entanglement physics is observed \cite{PSCR2012Likhtman, PRL1991DueringKremer, M2014KalathiKumar, PNAS2015ZhangWolynes}. To \textit{dynamically} destroy entanglements, typically the interchain repulsion energy between a pair of beads is (unphysically) made finite at full overlap ($r=0$) and/or the bonding spring constant is softened. Different simulation studies \cite{PSCR2012Likhtman, PRL1991DueringKremer, M2014KalathiKumar, PNAS2015ZhangWolynes} soften the interchain repulsion, soften the intrachain bonding potential, or do both, and to various degrees and in technically different ways. In practice, this exercise has a stochastic character due to the lack of a fundamental guiding principle for how to ``turn off'' entanglements. The generic aspects of the various implementations effectively introduce a modest finite barrier, often cited \cite{PRL1991DueringKremer, M2014KalathiKumar} as $\sim 2-5 k_BT$, for bond crossing events to occur frequently enough to destroy the objective metrics of entanglement dynamics. A definitive answer to the question of how much crossing is necessary to destroy entanglement dynamics, as a function of observation time and $N$, has not been adequately addressed. We are aware of only one systematic attempt \cite{ChangYethiraj}.
 
\subsection{Dynamical Chain Crossing in the Theory}
The theoretical question is can the dynamic constraints that enter our formulation be rationally softened to make ``entanglement localization go away''? We study, in two distinct ways, the role of connectivity and uncrossability in the context of the full $N$ order parameter dynamic theory. First, to explicitly allow for local chain crossing, the input liquid pair structure from thread PRISM theory can be varied. This is done by relaxing the ``hard'' no-overlap condition in Eq. (32) such that the contact pair distribution function is non-zero, $g_0 \neq0$. This means segments on different chains can sit on top of each other, and in the context of our theory implies dynamic chain crossing occurs. The analysis of Sec. IV.A can be repeated with this ``soft core condition'' to find:
\begin{align}
\xi_p^{-1} = \frac{\pi}{3} \rho_s \sigma^2 (1-g_0) + \sqrt{\frac{12}{N\sigma^2}}
\label{eq49}
\end{align}
which implies the dynamic coupling constant in Eq.(36) is weakened:
\begin{align}
\rho_s C_0^2 S_0 = - C_0 \rightarrow -C_0 (1-g_0)^2
\label{eq50}
\end{align}

 	The dotted and dashed curves in Figure 3 show the pair distribution function for $g_0 = 0.15$. The primary effect of this softening is to modify $g(r)$ on the local scale where the force acts and dynamic contacts are present in the theory; there is essentially no effect on the correlation hole. As an alternative perspective, $g_0$ can be roughly interpreted in terms of an interchain, segment-segment, potential-of-mean-force, and two segments can now overlap at a non-infinite (it is infinite for $g_0 = 0$) energy cost of $-k_BT \log(g_0)$. The latter is, for example, $\approx 2k_BT$ for $g_0 = 0.15$. Thus, as $g_0$ increases, chains on average cross more, which mimics within our theory changing Eq. (47) in MD simulations. 

	A second approach to allowing chain crossability is to soften the bonding entropic spring constant in Eq.(9) as:
\begin{align}
k_s = \alpha \frac{3k_BT}{\sigma^2}
\label{eq51}
\end{align}
where $\alpha = 1$ corresponds to the standard Gaussian chain and $\alpha < 1$ to a chain with larger bond length \textit{fluctuations}. When we implement this idea, the equilibrium packing structure is \textit{not} changed, which isolates the effects of intramolecular softening on our \textit{dynamical} predictions, and is the analog of changing Eq. (48) in simulations. It allows the study of how local aspects of connectivity affect entanglement localization. 

	Figure 11 shows the results of our analysis. In the inset, calculations of the tube diameter normalized by $2R_g$ for $N=5000$ are shown. The dotted curves consider the first approach where the bonding spring constants are fixed and local crossability is varied via $g_0$ (shown on the x-axis). The furthest right dotted curve shows results for $\alpha = 1$. As overlap is allowed ($g_0$ increases) the tube diameter swells until at (the intuitive) value of $d_T \approx 2R_g$, where $g_0 = 0.29$, entanglement localization is predicted to disappear in an abrupt manner. Analogous calculations with weakened bonding springs of $\alpha = 0.9$ and $0.8$ are also shown. As $\alpha$ decreases, similar behavior is predicted as found for $\alpha = 1$, but, sensibly, less interchain overlap or crossing (smaller $g_0$) is required to destroy mesoscopic localization since bonded spring softening further enhances crossing. Importantly, for all cases we find $d_T \approx 2R_g$ when entanglement localization is destroyed, consistent with our findings in Section V. Moreover, the prediction that the destruction of entanglement localization is a \textit{discontinuous} transition provides a theoretical basis for the phenomenological notion of a well-defined ``entanglement crossover'' value of $N=N_e$. 

\begin{figure}[b!]
\hspace{-0.35 in}
 	\includegraphics[height = 2.75 in]{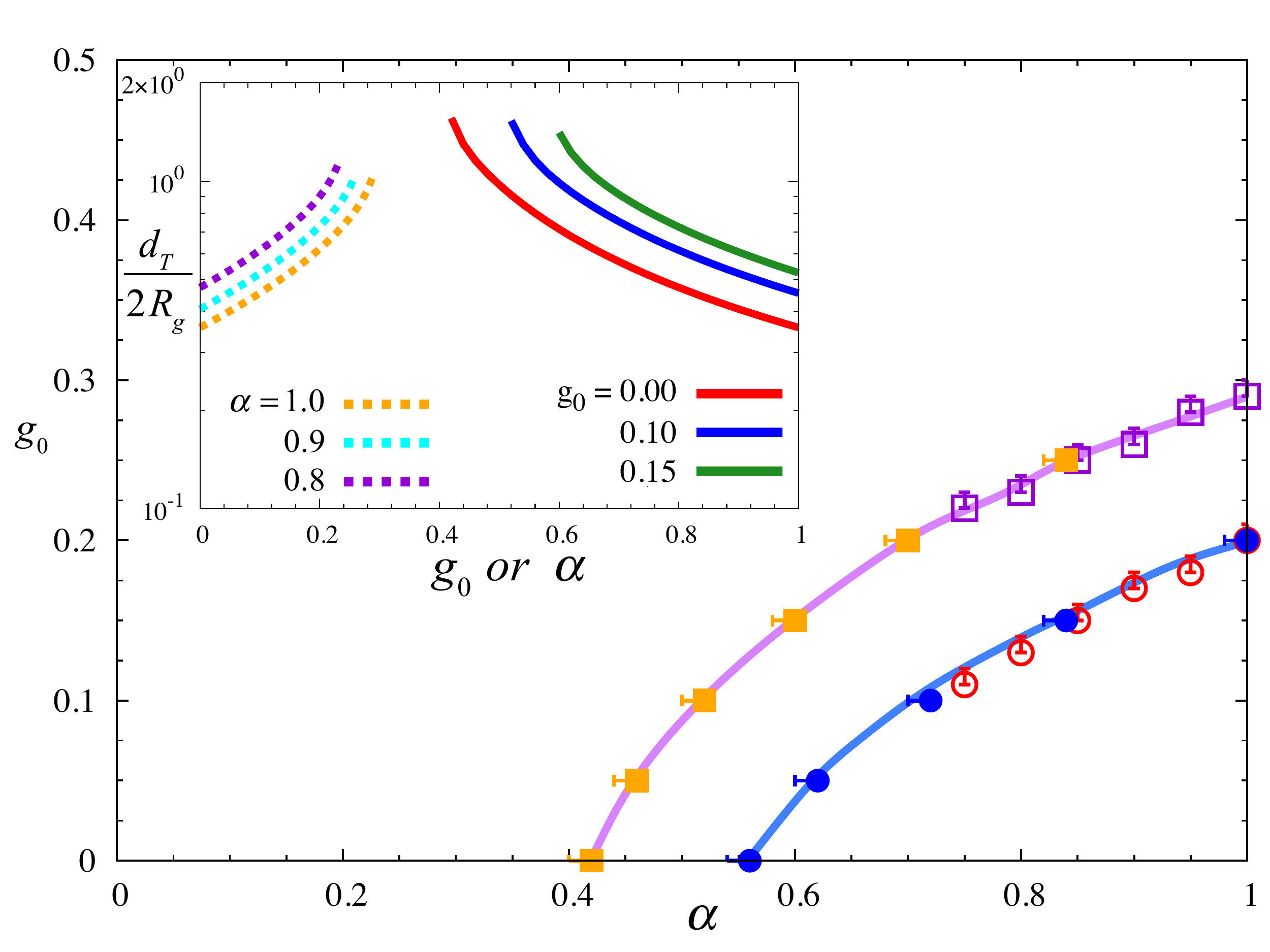}%
	\caption{(main) Effect of chain crossability for $\sigma/p=2$. (main) Disentanglement phase diagram, where $g_0$ is pair distribution function at contact and $\alpha \equiv k_s \sigma^2/3k_BT$ is the reduced Rouse spring constant. The bottom right corner corresponds to the mesoscopic localized regime. Results are shown for $N=5000$ (squares) and $N=3000$ (circles). Closed (open) symbols are numerical results generated from horizontal (vertical) cuts in the phase diagram. (inset) The tube diameter normalized by the diameter of gyration, $d_T/2R_g$, for $N=5000$. The dashed curves on the left are at constant $\alpha$ where $g_0$ is varied (x-axis), the solid curves on the right are at fixed $g_0$ and $\alpha$ is varied (x-axis).\label{fig11}}
\end{figure}

	Consequences of softening entropic springs are also studied in more detail at fixed $g_0$ and $N=5000$. The results are shown by the right most, solid curves in the inset of Fig. 11 for three values of $g_0$. The left most curve shows results for an infinite interchain repulsion ($g_0=0$), while the other two curves implement softening. In all cases, mesoscopic localization weakens as bonding springs are softened until it is completely destroyed when $d_T \approx (2-3) R_g$. When both bonding springs and interchain contact repulsion are softened, mesoscopic localization is more easily destroyed. 

	By systematically investigating the two ways of allowing chain crossing one can construct a ``disentanglement phase diagram''. Of course, in simulation such an idealized crisp transition will be blurred to some extent. The main frame in Fig. 11 shows results for typical chain lengths simulated: $N=5000$ (outer curve, $N/N_e \approx 13$ in the uncrossable limit) and $N= 3000$ (inner curve, $N/N_e \approx 7$). In the bottom right corner mesoscopic localization exists, while above the curves there is none. Clearly, entanglement localization is predicted to be a fragile phenomenon since it can be destroyed by introducing modest chain crossing. As expected, it is easier to destroy mesoscopic localization for shorter chains, i.e., less softening is required. 

Overall, we believe our theory qualitatively agrees with the results of MD simulations that probe the effect of chain crossability on entanglement physics. Crucially, we have shown that, within our statistical mechanical framework, the theory ``knows'' that if chain crossing is dynamically allowed, entanglement localization can be destroyed. Segment-segment excluded volume potentials and bond length fluctuations must be carefully selected to prevent sufficient chain crossing and loss of entanglements. The long range correlation hole aspect of the equilibrium liquid structure is irrelevant. New simulation studies of chain crossing should be performed to more precisely test our ideas. 
 
\section{DISCUSSION}
We have developed a segment-scale, self-consistent, $N$ dynamic-order-parameter, force-based theory for the breakdown of the isotropic Rouse model due to the combined influences of interchain repulsions and chain connectivity. Mesoscopic entanglement localization and its consequences in the absence of ergodicity-restoring anisotropic motions have been predicted. The key quantity is the intermolecular force memory function matrix which captures arrested segmental correlations that are nonlocal in separation along a tagged chain. An effective local force (contact) description and a universal Gaussian thread model are adopted, and the theory is closed at the chain dynamic second moment matrix level. The theoretical centerpiece (per Figure 4) is emergent ``statistically persistent, 2-chain contacts'' for long enough polymers and/or high enough concentrations. This perspective seems consistent with simulation-based deductions that have identified the crucial role of such long-lived contacts  \cite{PSCR2012Likhtman,PR1996Ben-NaimGrest, *JCP1997SzamelWang,SM2014Likhtman, *M2014LikhtmanPonmurugan, PR2012Everaers, *M2014QinMilner, *M2012AnogiannakisTzoumanekas}.
	
For long chain melts, the emergent entanglement localization and entropic elasticity predictions are in near quantitative agreement with experiments. The onset of mesoscopic localization in solution scales as the semi-dilute crossover with a sensible numerical prefactor. Due to the $N$-order parameter nature of the theory, various off-diagonal dynamic properties (chain second moment, force correlations in segment and mode space, coherent structure factor) were calculated which provide deeper insight into the spatial structure of the localized polymer conformational state. 
	
The role of chain uncrossability and connectivity on mesoscopic localization was investigated. As the local crossability and/or the intrachain entropic spring softness are increased, the tube diameter grows until $d_T \approx 2R_g$ at which point localization vanishes in a discontinuous manner. A full dynamic phase diagram for entanglement localization destruction was constructed, which appears to be qualitatively consistent with (limited) simulation studies. Testable predictions are made. This agreement reinforces our proposal that local uncrossability and chain connectivity, in concert with a self-consistent treatment of polymer motion and constraining forces, lead to the breakdown of isotropic Rouse theory and the emergence of mesoscopic localization. 
	
 To the best of our knowledge, we have constructed the first segment-scale, force level theory for the failure of the isotropic Rouse model and emergence of kinetic arrest and entanglement localization. We believe this is a significant step towards a full understanding of the emergence and spatial nature of entanglement phenomena. It might be useful as input to coarse grained stochastic models based on slip-links or slip-springs  \cite{M2005Likhtman,*M2008ReadJagannathan, ARCBE2014SchieberAndreev, *M2006NairSchieber, *PRL2008KhaliullinSchieber}. But we also believe a unified, segment scale, microscopic theory for tube localization, entanglement emergence, \textit{and} anisotropic long time reptation for flexible coils remains to be created. A deeper study of the role of a priori using the Gaussian thread model as input to the dynamical theory, versus it naturally emerging in a dynamic treatment that automatically does the proper structural self-averaging, is an open issue in statistical mechanics. But we believe that the mesoscopic nature of entanglement localization physically justifies using the thread polymer model of liquid structure. 
 
	The present version of the theory can be applied to other macromolecular problems where connectivity and interchain forces are important. Most fundamentally, the role of spatial dimensionality is of interest. Are there upper and lower critical dimensions for which random coil melts lose entanglement localization at any value of $N$? Is our prediction of a mesoscopic ``topological glass transition'' relevant to dense liquids of ring polymers and biological analogs \cite{RPP2014HalversonSmrek, *JCP2010BohnHeermann, EL2013LoTurner, *PNAS2016MichielettoTurner}? More generally, if macromolecules are not ideal random walks but still interpenetrating fractals, how does the breakdown of the isotropic Rouse model and entanglement localization evolve as a function of mass fractal and spatial dimensions? Systems and questions of direct experimental interest can also be studied. By adding an external force to the GLEs, one can address how mechanical stress or strain may destroy entanglement localization, a topic of high interest in nonlinear rheology  \cite{SM2015Wang, *JPCM2015SnijkersPasquino, *ML2015FalzoneRobertson-Anderson}. The role of chemical crosslinking can be treated by changing the boundary conditions on the GLEs. Finally, the interplay between chain connectivity and attractive (for gel-forming homopolymers, copolymers, ionomers) and/or long range repulsive (charged polymers) forces on transient localization and elasticity of unentangled polymers can be investigated.

\appendix
\section{General Theory Formulation}
\subsection{Arrested Force-Force Time Correlation Function Matrix} 
	The foundation of our dynamical theory is the formally intractable force memory function matrix of Eq.(10) which involves 4-point space-time correlations associated with two segments on the tagged chain and two segments from chains in the surrounding matrix. To determine it,  we consistently invoke a nonperturbative, but Gaussian, self-consistent density field approximation. For structure, this means a Gaussian single chain model and the thread version of PRISM theory \cite{ACP1997SchweizerCurro,M1988SchweizerCurro} for packing correlations. For dynamics, it implies a Gaussian-like treatment of time-dependent force correlations between the tagged chain and its surroundings which involves three key simplifications. (i) Dynamic self-consistency via replacing projected dynamics in Mori-Zwanzig theory  \cite{JCP1974Zwanzig, 2001Zwanzig} by true dynamics, $e^{QLt} \rightarrow e^{Lt}$. (ii) The segment-segment (or site-site) interchain forces between chains are replaced by the gradient of an effective pair interaction (in the spirit of integral equation theory \cite{ACP1997SchweizerCurro,M1988SchweizerCurro}, the solvated electron problem \cite{JCP1984ChandlerSingh}, and mode coupling theory \cite{JSMTE2005ReichmanCharbonneau, 2008Gotze} for spheres), corresponding to:
\begin{align}
\vec{F}_\alpha = \sum_{j, \gamma} k_BT \vec{\nabla} C_{\alpha m}\left(\left|\vec{r}_\alpha - \vec{r}_{j, \gamma} \right|\right)\label{eqA1}
\end{align}
where $C(r)$ is the short range (for neutral polymers) site-site direct correlation function and $\vec{F}_\alpha$ is the total effective force on segment $\alpha$ of the tagged chain due to all surrounding polymers (sum over $j$) each of which is composed of $N$ segments (sum over $\gamma$). (iii) All higher than pair correlation functions are factorized per the Gaussian idea. 
	
To implement the above strategy, recall the force-force time correlation matrix:
\begin{align}
K_{\alpha \beta}(t) = \frac{\beta}{3} \left< \vec{F}_\alpha (t) \cdot \vec{F}_\beta (0)\right>
\label{eqA2}
\end{align}
In the spirit of the recently formulated Òprojectionless dynamics theoryÓ \cite{PRL2015DellSchweizer}, this can be written in a density field representation as:
\begin{align}
\vec{F}_\alpha(t) = \int d\vec{r} \int d\vec{r}\,' \rho_\alpha (\vec{r}, t) \vec{F}( \vec{r} - \vec{r} \,') \rho_m (\vec{r} \,', t)
\label{eqA3}
\end{align}
where the single tagged segment and collective matrix density fields are, respectively,
\begin{align}
\rho_\alpha(\vec{r}, t) & \equiv \delta(\vec{r} - \vec{R}_\alpha(t)) \nonumber \\
\rho_m (\vec{r}\,', t) & \equiv \sum_{j, \gamma} \delta(\vec{r}\, ' - \vec{R}_{j, \gamma}(t))
\label{eqA4}
\end{align}
Substituting Eq. (A3) in Eq. (A2) yields:
\begin{align}
K_{\alpha \gamma}(t) &= \frac{\beta}{3} \int d\vec{r} \int d\vec{r} \,' \int d\vec{r}\, '' \int d\vec{r}\,'''\vec{F}_\alpha (\vec{r} - \vec{r} \,') \nonumber \\
&\cdot \vec{F}_\gamma ( \vec{r} \, '' - \vec{r} \, ''')\left<\rho_\alpha(\vec{r}) \rho_m(\vec{r}\,') \rho_\beta(\vec{r} \, '', t) \rho_m(\vec{r} \, ''', t) \right> 
\label{eqA5}
\end{align}
Invoking simplification (iii) above implies:
\begin{flalign}
\left<\rho_\alpha(\vec{r}) \rho_m(\vec{r}\,') \rho_\beta(\vec{r} \, '', t) \rho_m(\vec{r} \, ''', t) \right> \nonumber \\
& \hspace{-1.2 in} \approx  \left<\rho_\alpha(\vec{r})\rho_\beta(\vec{r} \, '', t) \right>\left< \rho_m(\vec{r}\,') \rho_m(\vec{r} \, ''', t) \right> \nonumber \\
&\hspace{-1.2 in} = \omega_{\alpha \gamma}(\vec{r} - \vec{r} \, '', t) \rho_s S(\vec{r} \,' - \vec{r} \, ''', t)
\label{eqA6}
\end{flalign}
where the intrachain dynamic structure factor matrix and the time dependent collective density fluctuation structure factor are given by the first and second terms in the second line, respectively. Using Eq. (A6) in Eq. (A5), simplification (ii), global isotropy, and the Fourier convolution theorem, one obtains the final expression of Eq.(17):
\begin{align}
K_{\alpha \beta}(t) = \frac{\beta^{-1} \rho_s}{3} \int \frac{d\vec{k}}{(2\pi)^3} \left( kC(k) \right)^2 \omega_{\alpha\beta}(k,t) S(k,t). 
\label{eqA7}
\end{align}

\subsection{Kinetically Arrested Rouse Amplitudes}
	The long time limit solution for the (orthogonal) Rouse mode correlation functions $\left< \vec{X}_p (\infty) \cdot \vec{X}_p(0)\right>$  and $\left< X_p^2(\infty)\right>$ follow from Eq. (13) as: 
\begin{align}
\hspace{-0.5 in}
\zeta_s \frac{d}{dt} \left< \vec{X}_p (t) \cdot \vec{X}_p(0)\right> &= - \kappa_p \left< \vec{X}_p (t) \cdot \vec{X}_p(0)\right>  + K_p(\infty) \left<X_p^2(0)\right> 
\label{eqA8} \\
\hspace{-0.25 in}
\zeta_s \frac{d}{dt} \left< X_p^2(t)\right> = - 2\kappa_p& \left<X_p^2(t)\right>  + 2K_p(\infty) \left<\vec{X}_p (t) \cdot \vec{X}_p(0)\right> \nonumber \\
+6k_BT
\label{eqA9}
\end{align}
where the total spring constant is $\kappa_p \equiv \lambda_p +K_p(\infty)$. In the kinetically arrested limit, the time derivatives on the left hand side of the equations vanish. Solving the resultant two coupled algebraic equations, with the initial condition $\left< X_p^2(0)\right>= 3k_BT/\lambda_p$ , yields:
\begin{align}
 \left< \vec{X}_p (\infty) \cdot \vec{X}_p(0)\right> = \frac{3k_BT}{\lambda_p + K_p(\infty)} \left[ \frac{K_p(\infty)}{\lambda_p}\right]
\label{eqA10} \\
\left<X_p^2(\infty)\right> = \frac{3k_BT}{\lambda_p + K_p(\infty)} \left[ 1+ \frac{K_p(\infty)}{\lambda_p + K_p(\infty)}\frac{K_p(\infty)}{\lambda_p}\right]
\tag{A11}
\label{eqA11}
\end{align}
\section{Limiting Cases of the General Dynamic Theory}
	\subsection{Glass Localization and Diagonal Memory Function Limit}
	If only the diagonal ($\alpha = \gamma$) force memory term is retained in Eq. (17) then we find sub-segmental scale glassy localization ($r_L << \sigma$) is predicted. This is not of present interest, and the coarse-grained Gaussian thread model is not really appropriate on the local scales on which glassy dynamics occurs. It is for these reasons that in our analysis of mesoscopic localization the diagonal term in the memory function matrix is dropped. Nevertheless, it is instructive to sketch the diagonal limit of the theory, which is given by:
\begin{align}
K_{\alpha \beta}(\infty) & \approx K_{\alpha \alpha}(\infty) \cdot \delta_{\alpha \gamma} \equiv K_{Diag} \delta_{\alpha \gamma} \nonumber \\
 K_{Diag}(\infty) & =  \frac{\beta^{-1} \rho_s}{3} \int \frac{d\vec{k}}{(2\pi)^3} \left( kC(k) \right)^2  S(k) \exp\left(- k^2r_L^2/3\right) 
\tag{A12}
\label{eqA12}
\end{align}
where the segment localization length $r_L^2 \equiv \delta \mu_{\alpha \alpha}^{(2)} (\infty) = \left< \left(\vec{R}_\alpha(\infty) -\vec{R}_\alpha(0) \right)^2 \right>$. This limit applies to the full matrix memory function theory \textit{if} the dynamic localization length is much smaller than the segment size, whence $K_{Diag}(\infty) >> 3k_BT/\sigma^2$. Exploiting these simplifications, Eq.(26) yields a single self-consistent relation for the localization length:  
\begin{align}
r_L^2 = \frac{3k_BT}{K_{Diag}(\infty)}
\tag{A13}
\label{eqA13}
\end{align}

	 If one adopts a freely jointed chain model (bond length $\sigma$) with \textit{non-zero} hard-sphere site-site excluded volume (site hard core diameter $d = 3 \sigma/4$), and standard numerical PRISM theory \cite{ACP1997SchweizerCurro} for the required structural input, the local cage packing correlations are captured. We find that tight ($r_L << \sigma$) glassy localization is predicted which emerges at a high space-filling volume fraction $\eta \equiv \pi \rho_s d^3/6 \approx 0.42$, which is almost identical to what is predicted (at the single particle dynamics level of theory) for a liquid of disconnected hard spheres \cite{JCP2003SchweizerSaltzman}. Moreover, it corresponds to a condition on the packing length of $\sigma/p \gtrsim 1.9$, which is well into the dense polymer melt regime \cite{M1994FettersLohse}. Finally, and most crucially, we emphasize that, consistent with physical intuition, we  find that the diagonal theory \textit{never} predicts mesoscopic localization.  

\subsection{Center-of-Mass Theory}
	Here we derive the center-of-mass (COM) force-force correlations from the full matrix theory. The starting point is Eq. (12) where $\Delta \alpha \equiv \left| \alpha - \gamma \right|$. The COM theory is a one dynamic order parameter description in the long wavelength limit where $K_p(\infty) \approx K_0(\infty)$. Recalling $\psi_0(\Delta \alpha) = A_0 = \sqrt{1/N}$, one obtains from Eq. (17):
\begin{align}
 K_{0}(t) & =  \frac{\beta^{-1} \rho_s}{3} \int \frac{d\vec{k}}{(2\pi)^3} \left( kC(k) \right)^2  S(k, t) \sum_{\Delta \alpha = 0}^N \omega_{\alpha \gamma}(k,t) 
 \tag{A14}
\label{eqA14}
\end{align}
Since $\omega_{\alpha \gamma}$ is only a function of $\left|\alpha - \gamma\right|$ , the final term in Eq. (A14) can be rewritten as:
\begin{align}
 \sum_{\Delta \alpha = 0}^N \omega_{\alpha \gamma}(k,t) &= \frac{1}{N} \sum_{\gamma = 0}^N  \sum_{\Delta \alpha = 0}^N \omega_{\alpha \gamma}(k,t)  \nonumber \\
 &\approx \frac{1}{N}  \sum_{\alpha, \gamma = 1}^N \omega_{\alpha \gamma}(k,t) \equiv \omega(k,t)
 \tag{A15}
\label{eqA15}
\end{align}
where the approximate equality is exact for long chains. The COM limit of the dynamic theory of Eq.(27) then follows. 

\section{Analytic Results in Gaussian Thread Polymer Limit}
We derive several analytic results cited in the text for the universal description of a polymer chain and liquid structure, the Gaussian thread model.

\subsection{Simplified Force Memory Function}  
The memory function matrix in the polymer thread limit where $C(k)=C_0$ is:
\begin{align}
 K_{\alpha \gamma}(t) & =  \frac{\beta^{-1} \rho_s}{3} C_0^2 \int \frac{d\vec{k}}{(2\pi)^3} \left[\vec{k}\, \omega_{\alpha \gamma}(k,t) \right] \cdot \left[ \vec{k} S(k, t)\right] 
 \tag{A16}
\label{eqA16}
\end{align}
Using the definition of the packing length, and re-writing the Fourier integration in real space using ParsevalÕs theorem, one finds:
\begin{align}
 K_{\alpha \gamma}(t) & =  \frac{k_BT}{\sigma^2} \frac{C_0^2}{3p} \int d\vec{r} \left[\vec{\nabla} \omega_{\alpha \gamma}(\vec{r},t) \right] \cdot \left[ \vec{\nabla} S(\vec{r}, t)\right] 
 \tag{A17}
\label{eqA17}
\end{align}
To gain intuition concerning the arrested dynamical constraints on a tagged chain, this expression is analyzed at $t=0$. The density fluctuation structure factor is highly local in real space, decaying exponentially beyond the density correlation or mesh length $\xi_p$ (or packing length) which is small compared to the mesoscopic localization length. Thus, for simplicity, we replace the screened Yukawa form of Eq.(30) by a step function : 
\begin{align}
S(r) \approx \tilde{S} \Theta_- (r-\xi_p)
\tag{A18}
\label{eqA18}
\end{align}
where $\tilde{S} \propto \left(\xi_p/\sigma \right)^2 \propto S_0$  since it is an average of $S(r)$ inside the mesh length. Taking the \begin{align}
 K_{\alpha \gamma}(0) & \approx - \frac{k_BT}{\sigma^2} \frac{4\pi C_0^2 \tilde{S}}{3p} \left[\frac{\partial}{\partial r} \omega_{\alpha \gamma}(\vec{r}) \right]_{r=\xi_p} 
 \tag{A19}
\label{eqA19}
\end{align}
This is consistent with Fig.4a which indicates the 4-body slow configuration involves two segments on the tagged chain initially within a density correlation length and segments on the two other chains that exert forces on them, thereby forming a ``tight contact''.
 
	The physical content of the derivative term in Eq.(A19) can be elucidated by writing its formal definition\cite{1986DoiEdwards}: 
\begin{align}
\frac{\partial}{\partial r} \omega_{\alpha \gamma}(\vec{r}) &= \frac{\partial}{\partial r} \left < \delta \left( \vec{r} - \vec{R}_\alpha + \vec{R}_\gamma \right)\right >  \nonumber \\
& = \int D\vec{R}_\alpha \Psi \left( \vec{R}_\alpha \right) \frac{\partial}{\partial r} \delta \left( \vec{r} - \vec{R}_\alpha + \vec{R}_\gamma \right)
\tag{A20}
\label{eqA20}
\end{align}
where $D\vec{R}_\alpha$ indicates an integral over all possible chain conformations and $\Psi\left(\vec{R}_\alpha \right)$ is the conformational distribution function. Using the delta function, the derivative can be re-expressed as $\partial/\partial\vec{r} = \partial/\partial\vec{R}_\alpha - \partial/ \partial \vec{R}_\gamma$. Integration by parts then yields:
\begin{align}
\frac{\partial}{\partial r} \omega_{\alpha \gamma}(\vec{r}) & = \int D\vec{R}_\alpha \, \delta \left( \vec{r} - \vec{R}_\alpha + \vec{R}_\gamma \right) \nonumber \\
& \qquad \times \left( \frac{\partial}{\partial \vec{R}_\gamma } -  \frac{\partial}{\partial \vec{R}_\alpha}\right)\Psi \left( \vec{R}_\alpha \right) 
\tag{A21}
\label{eqA21}
\end{align}
Finally, in terms of the polymer free energy, $A$ , one has $\Psi\left(\vec{R}_\alpha\right) = Z \exp \left[ -\beta A\left(\vec{R}_\alpha\right)\right]$, where for a Gaussian chain $A\left(\vec{R}_\alpha\right) = \frac{1}{2} k_s \sum_{\alpha = 1}^N \left| \vec{R}_\alpha - \vec{R}_{\alpha - 1} \right|^2$ \cite{1986DoiEdwards}. From this one finds:
\begin{align}
\frac{\partial}{\partial r} \omega_{\alpha \gamma}(\vec{r}) & = \left< \delta \left( \vec{r} - \vec{R}_\alpha + \vec{R}_\gamma \right) \left( \frac{\partial A}{\partial \vec{r}_\gamma } -  \frac{\partial A}{\partial \vec{r}_\alpha}\right) \right>
\tag{A22}
\label{eqA22}
\end{align}
In conjunction with Eq.(A19), this result implies the key quantity is the difference in the intrachain spring force on the two tagged segments held a distance $\xi_p$ apart. Geometrically, this is related to a conformational ``loop'' between tagged segments as sketched in Fig. 4a. In the long time localized state, these force correlations do not fully relax, implying some of the initial conformational information is retained. 

\subsection{Tube Diameter in Center-of-Mass Theory}
         	Here we derive the analytic scaling of the tube diameter in the COM theory of Eq. (38). Inserting the thread PRISM structural relations in Eq.(27) yields:
\begin{align}
\hspace{-0.0in}
K_{COM}(\infty) &= \frac{k_BT}{6 \pi^2 p \sigma^2} \int_0^\infty dk\, k^4 C_0^2 \frac{S_0}{1+k^2 \xi_p^2} \nonumber \\
& \qquad \qquad \times  \frac{1}{N^{-1}+k^2 \sigma^2/12}\exp\left(- \frac{k^2d_T^2}{12} \right) 
\tag{A23}
\label{eqA23}
\end{align}
Given the mesoscopic nature of localization, $\sigma << d_T < 2R_g$, the wavelength dependence of the static structure factor is irrelevant ($k\xi_p << 1$), and thus 
\begin{align}
\hspace{-0.0in}
K_{COM}(\infty) &= \frac{k_BT}{6 \pi^2 p \sigma^2} C_0^2 S_0 \int_0^\infty dk\, \frac{k^4}{N^{-1}+k^2 \sigma^2/12} \nonumber \\
& \qquad \qquad \qquad \qquad \qquad \times \exp\left(- \frac{k^2d_T^2}{12} \right) 
\tag{A24}
\label{eqA24}
\end{align}
Importantly, all structural information reduces to a single ``coupling constant'' 
\begin{align}
C_0^2 S_0 = &= \frac{\pi^2 \sigma^6}{108} \left[ \frac{1+(6p/\pi \sigma) \sqrt{12/N}}{1+(3p/\pi \sigma) \sqrt{12/N}}\right]^2
\tag{A25}
\label{eqA25}
\end{align}
In the short (long) chain limit, one has the intrinsic result $C_0^2S_0 = \pi^2 \sigma^6/27$ $( = \pi^2 \sigma^6/108)$. The integral in Eq. (A24) is analytically performed yielding: 	
\begin{align}
K_{COM}(\infty) &= \frac{12 \sqrt{3\pi} k_BT}{\pi^2 p d_T^3 \sigma^4} C_0^2 S_0 \left[ 1 - \frac{d_T^2}{3R_g^2}  \right . \nonumber \\
& \left .+ \frac{1}{3} \sqrt{\frac{\pi}{6}} \frac{d_T^3}{R_g^3} e^{d_T^2/6R_g^2} \left( 1 - \mbox{erf}\left[\frac{d_T}{\sqrt{6} R_g} \right]\right) \right]
\tag{A26}
\label{eqA26}
\end{align}
where $\mbox{erf}(x)$ is the error function. Combining this with the self-consistency Eq.(26) gives the closed COM theory for the tube diameter, which can be numerically solved. 

To gain further insight, two limits are considered: chains near the onset of localization and $N>>1$. Near onset, we expect (and predict from the full theory) that $d_T \approx 2R_g$. Then all terms inside the square brackets in Eq.(A26) are of order unity, and here we set them literally to unity for simplicity. Combining this with the long chain limit coupling constant gives:
\begin{align}
K_{COM}(\infty) &= k_BT \frac{\sqrt{3\pi}}{9} \frac{\sigma^2}{pd_T^3} 
\tag{A27}
\label{eqA27}
\end{align}
The localization length self-consistency equation is:
\begin{align}
d_T^2 &=  \frac{6k_BT}{K_0(\infty)} \frac{1}{N} + \frac{4 \sigma^2}{\pi} \left( \frac{3k_BT}{2K_0(\infty) \sigma^2}\right)^{1/2} \nonumber \\
& \qquad \qquad \qquad  \times  \arctan \left[\pi \left( \frac{3k_BT}{2K_0(\infty) \sigma^2}\right)^{1/2}\right]
\tag{A28}
\label{eqA28}
\end{align}
where the first term comes from the COM mode and the second term comes from the many polymer internal degrees of freedom. Given $d_T \approx 2R_g$, the argument of the arctangent is large, and thus $\arctan(x\rightarrow \infty) \approx \pi/2$. Substituting Eq.(A27) in Eq.(A28), and dividing through by $d_T^2$, one finds a simplified self-consistency relation: 
\begin{align}
1 &=  \frac{54}{(3\pi)^{1/2}} \frac{1}{N} \frac{p}{\sigma} \frac{d_t}{\sigma}  + \frac{54^{1/2}}{(3\pi)^{1/4}} \left( \frac{p}{\sigma} \right)^{1/2} \left(\frac{\sigma}{d_T}\right)^{1/2}
\tag{A29}
\label{eqA29}
\end{align}
When $d_T \rightarrow 0$ the second term on the right hand side (RHS) of this equation diverges, while if $d_T \rightarrow \infty$ then the first term on the RHS diverges. Thus, there must be a minimum of the RHS curve. \textit{But}, whether it solves the equation (i.e., RHS=1) is \textit{not} guaranteed, and depends on the value of $N$ and packing length in units of the statistical segment length. This is the mathematical origin of our prediction of a discontinuous mesoscopic localization transition. The value of the tube diameter at the RHS minimum is: 
\begin{align}
\frac{d_{T, on}}{\sigma}  &=  \frac{\pi^{1/6}}{2\cdot3^{5/6}}N^{2/3} \left(\frac{\sigma}{p}\right)^{1/3}  
\tag{A30}
\label{eqA30}
\end{align}
Substituting this into Eq. (A29) and solving for the onset inverse packing length yields:	
\begin{align}
\frac{\sigma}{p_{on}}  &=  \frac{81}{\pi^{1/2}}N^{-1/2} \approx 13 \rho_s^* \sigma^3  
\tag{A31}
\label{eqA31}
\end{align}
where $\rho_s^*$ is the semi-dilute crossover. Eq.(A31) can be compared to the numerical COM calculation that found $\sigma/p_{on} \approx19 \rho_s^* \sigma^3$(Fig.10 inset). 

Equation (A31) can alternatively be interpreted by fixing the packing length and considering the onset chain length. This yields that for small enough chain lengths Eq.(A29) does not have a localized solution. However at $N_{on} = \pi (\sigma/81p)^2$, corresponding to a critical number of internal conformational modes, a solution emerges indicating mesoscopic localization. This establishes that the emergence of entanglement localization is due to the interplay between the COM and internal degrees of freedom.
	
Finally, in the $N\rightarrow \infty$ limit the self-consistent localization equation always has a solution. The square bracket in Eq.(A26) is again set to unity. Following the same analysis as above, the intrinsic ($N$-independent) tube diameter of Eq.(38) is derived: 
\begin{align}
d_T  &= \left( \frac{54p}{\sqrt{3\pi}}\right) 
\tag{A32}
\label{eqA32}
\end{align}
\subsection{Scaling of Matrix Dynamic Force Correlations}
	Here we derive the large separation scaling of the matrix dynamic force correlations of Eq.(42).The starting point based on the Gaussian thread liquid structure is:
\begin{align}
\hspace{-0.0in}
K_{\alpha \gamma}(\infty) & = \frac{k_BT}{6 \pi^2 p \sigma^2} \int_0^\infty dk\, k^4 C_0^2 \frac{S_0}{1+k^2 \xi_p^2}\exp\left[- \frac{k^2\sigma^2 \left| \alpha - \gamma \right|}{6} \right] \nonumber \\
& \qquad \qquad \times  \exp\left[- \frac{k^2\left(d_T^2/4 + \delta \mu_{\alpha \gamma}^{(2)}\right)}{6} \right] 
\tag{A33}
\label{eqA33}
\end{align}
Using $C_0^2S_0 = \pi^2 \sigma^6/108$ and $1+k^2 \xi_p^2 \approx 1$, valid for mesoscopic localization, one has:
\begin{align}
\hspace{-0.0in}
K_{\alpha \gamma}(\infty) & = \frac{k_BT \sigma^4}{648 p} \int_0^\infty dk\, k^4 \exp\left[- \frac{k^2\sigma^2 \left| \alpha - \gamma \right|}{6} \right] \nonumber \\
& \qquad \qquad \times  \exp\left[- \frac{k^2\left(d_T^2/4 + \delta \mu_{\alpha \gamma}^{(2)}\right)}{6} \right] 
\tag{A34}
\label{eqA34}
\end{align}
For large separations, $\left| \alpha - \gamma\right| >> d_T^2/\sigma^2$, the relevant wavevectors are small compared to the tube diameter. Hence, the first exponential in Eq.(A34) is dominated by the diagonal limit $\delta \mu_{\alpha \gamma}^{(2)} \approx\delta \mu_{\alpha \alpha}^{(2)} = d_T^2/4$, thereby yielding: 
\begin{align}
K_{\alpha \gamma}(\infty) & = \frac{k_BT \sigma^4}{648 p} \int_0^\infty dk\, k^4 \exp\left[- \frac{k^2\sigma^2 \left| \alpha - \gamma \right|}{6} \right] \nonumber \\
& \qquad \qquad \qquad \qquad \times \exp\left[- \frac{k^2 d_T^2}{12} \right] 
\tag{A35}
\label{eqA35}
\end{align}
Performing the Gaussian integration gives:
\begin{align}
K_{\alpha \gamma}(\infty) & = \sqrt{\frac{\pi}{12}} \frac{k_BT \sigma^4}{p d_T^5} \left[ 1 + \frac{2 \left| \alpha - \gamma \right | \sigma^2}{d_T^2}\right]^{-5/2}
\tag{A36}
\label{eqA36}
\end{align}
Hence, per Eq.(42) and the numerical results in Fig.6, when the segment separation is large an inverse power decay of the arrested force correlations is predicted:
\begin{align}
K_{\alpha \gamma}(\infty) & = K_{\alpha \alpha}(\infty) \left[ \frac{ \left| \alpha - \gamma \right | \sigma^2}{d_T^2}\right]^{-5/2} && \mbox{for } \left| \alpha - \gamma \right| > d_T^2/\sigma^2
\tag{A37}
\label{eqA37}
\end{align}
This power law is the combined consequence of structural homogeneity ($S(k)=S_0$) on large length scales and the long range correlations of a connected random walk chain as manifested via the Gaussian form of $\omega_{\alpha \gamma}(k)$ in Eq.(A35).  

Finally, comparing Eqs.(A36) and (A37) we find that the diagonal (zero separation) component of the memory function scales with the inverse packing length and tube diameter. Given for large chains $d_T \propto p$, this implies the following scaling relations:
\begin{align}
 K_{\alpha \alpha}(\infty)\propto p^{-1} d_T^{-5} \propto p^{-6} \propto d_T^{-6} 
 \tag{A38}
\label{eqA38}
\end{align}
These results have been verified in our numerical calculations in the long chain limit. \vspace{+0.1 in}

\begin{acknowledgments}
This work was supported by DOE-BES under Grant No. DE-FG02-07ER46471 administered through the Frederick Seitz Materials Research Laboratory. KSS acknowledges insightful email correspondences a number of years ago with Grzegorz Szamel and Mike Cates. We thank Arun Yethiraj for sending a copy of ref. \cite{ChangYethiraj}.
\end{acknowledgments}

\bibliography{Refs}

\end{document}